\documentclass[preprint,3p,times]{elsarticle}
\usepackage[activate={true,nocompatibility},final,tracking=true,kerning=true,spacing=true,factor=1100,stretch=10,shrink=10]{microtype}
\microtypecontext{spacing=nonfrench}
\usepackage[T1]{fontenc}
\usepackage{graphicx}
\usepackage{dcolumn}
\usepackage{amsmath}
\usepackage{bm}
\usepackage{hyperref}
\usepackage{subcaption}
\usepackage{siunitx}
\usepackage[capitalize]{cleveref}
\usepackage[english]{babel}
\usepackage{booktabs}
\usepackage[section]{placeins}

\DeclareMathOperator{\sgn}{\mathrm{sgn}}

\renewcommand{\bar}[1]{\mkern3.0mu\overline{\mkern-3.0mu#1}\mkern1mu}

\journal{Nuclear Physics B}

\begin{document}

\begin{frontmatter}

    \title{Universal non-equilibrium scaling of cumulants across a critical point}

    \author[JLU]{Leon J. Sieke\corref{cor1}}
    \ead{leon.j.sieke@physik.uni-giessen.de}
    \cortext[cor1]{Corresponding author}

    \author[BU]{Mattis Harhoff}
    \ead{mharhoff@physik.uni-bielefeld.de}

    \author[BU]{S\"oren Schlichting}
    \ead{sschlichting@physik.uni-bielefeld.de}

    \author[JLU,HFHF]{Lorenz von Smekal}
    \ead{lorenz.smekal@physik.uni-giessen.de}

    \affiliation[JLU]{organization={Institut f\"ur Theoretische Physik, Justus-Liebig-Universit\"at},
        addressline={Heinrich-Buff-Ring 16},
        city={35392 Gießen},
        country={Germany}}

    \affiliation[HFHF]{organization={Helmholtz Forschungsakademie Hessen f\"ur FAIR (HFHF)},
        addressline={Campus Gießen},
        city={35392 Gießen},
        country={Germany}}

    \affiliation[BU]{organization={Fakult\"at f\"ur Physik, Universit\"at Bielefeld},
        city={33615 Bielefeld},
        country={Germany}}

    \begin{abstract}
        We study the critical dynamics of a scalar field theory with \texorpdfstring{$Z_2$}{Z2} symmetry in the dynamic universality class of Model~A in two and three spatial dimensions with classical-statistical lattice simulations.
        In particular, we measure the non-equilibrium behavior of the system under a quench protocol in which the symmetry-breaking  external field is changed at a constant rate through the critical point.
        Using the well-established Kibble-Zurek scaling theory we compute non-equilibrium scaling functions of cumulants of the order parameter up to fourth order.
        Together with the static critical exponents and the dynamic critical exponent, these fully describe the universal non-equilibrium evolution of the system near the critical point.
        We further extend the analysis to include finite-size effects and observe good collapse of our data onto two-dimensional universal non-equilibrium and finite-size scaling functions.
    \end{abstract}


\end{frontmatter}

\tableofcontents

\section{Introduction}

In context of the search for the QCD critical point in heavy-ion collisions, a deep understanding of the out-of-equilibrium dynamics of the system is necessary to make well-grounded predictions for signatures in final states.
As the critical point is approached, the system will inevitably fall out of equilibrium as the relaxation time diverges, and evolve through a non-equilibrium regime before reaching an equilibrium state in the new phase or freezing out~\cite{Stephanov:1999zu,Berdnikov:1999ph}.
The critical fluctuations are in this case believed to be described by the dynamic universality class of Model~H in the classification scheme of Hohenberg and Halperin~\cite{Hohenberg:1977ym}, characterized by the diffusive nature of the fluctuations associated with the conserved entropy per baryon and the shear modes in the energy-momentum tensor~\cite{Son:2004iv}.

To describe the evolution of these fluctuations in heavy-ion collisions, various dynamical models have been developed, such as stochastic fluid dynamics~\cite{Bell:2007,Donev:2011,Usabiaga:2012,Murase:2016rhl,Hirano:2018diu,Nahrgang:2017oqp,Bluhm:2018plm,Singh:2018dpk,Chattopadhyay:2024jlh}, hydro-kinetics~\cite{Akamatsu:2016llw,Akamatsu:2018vjr,An:2019osr}, chiral fluid dynamics~\cite{Paech:2003fe,Nahrgang:2011mg,Nahrgang:2011vn,Herold:2013bi,Herold:2014zoa} and Hydro+~\cite{Stephanov:2017ghc,Rajagopal:2019xwg,Du:2020bxp}. For a comprehensive review, see~\cite{Bluhm:2020mpc}. However, numerical implementations of these models are often complex and calculations of fluctuating observables can be computationally expensive.
As a result, simpler dynamical models, such as the relaxational Model~A~\cite{Jiang:2017mji,Wu:2018twy,Jiang:2021fun} or diffusive Model~B~\cite{Sakaida:2017rtj,Nahrgang:2018afz,Nahrgang:2020yxm,Rougemont:2018ivt}, are often considered to study dynamical critical behavior across a wider range of phase transition scenarios.
Recent studies~\cite{Mukherjee:2015swa,Mukherjee:2016kyu} demonstrated the value of universal scaling functions describing the off-equilibrium evolution of higher order cumulants, as they are highly sensitive to the correlation length and thus interesting as signatures of a second-order phase transition~\cite{Stephanov:2008qz,Stephanov:2011pb,Nahrgang:2016ayr}.
Using the well-developed classical lattice model~\cite{Aarts:2001yx,Berges:2009jz,Schlichting:2019tbr,Schweitzer:2020noq,Schweitzer:2021iqk} naturally lends itself, as the changes needed to enable non-equilibrium studies are minimal, and boil down to dynamically changing two external parameters.

The dynamic scaling behavior of a system passing close by or through a critical point can be described by the Kibble-Zurek mechanism (KZM)~\cite{Kibble:1976sj,Zurek:1985qw,Zurek:1996sj}.
Originally describing the formation of topological defects in systems driven through continuous symmetry-breaking phase transitions, the KZM has since been generalized to a theory of finite-time scaling (FTS)~\cite{Gong:2010,Huang:2014oma,Feng:2016qmw} in analogy to the well established finite-size scaling (FSS) to describe non-equilibrium critical phenomena in a wide range of classical and quantum systems~\cite{delCampo:2013nla,Liu:2013nla,Dziarmaga:2009,Yin:2014}.
As the characteristic timescale of the system diverges at criticality, the system inevitably falls out of equilibrium, leading to the emergence of finite domains of local equilibrium in the new phase.
Depending on the specific dynamic equations governing the evolution, the system will exhibit different types of non-equilibrium behavior.
In the case of the relaxational dynamics of Model~A, it will slowly relax towards the new equilibrium state, from which on the evolution continues adiabatically.
The intermediate non-equilibrium regime will exhibit self-similar behavior controlled by universal scaling functions~\cite{Chandran:2012cjk}.
Knowing these scaling functions and the quench protocol controlling the whole process, one can exactly predict the long-distance behavior of the system.
This is true both for protocols that pass the critical point (trans-/cis-critical protocols), and those that asymptotically approach it (end-critical protocols).

In this work, we focus on trans-critical protocols (TCPs) taking the system across the critical point at a finite rate and aim to extract the universal scaling functions describing the off-equilibrium evolution of the order parameter and higher order cumulants under TCPs, where an external symmetry breaking field is changed at a constant rate through the critical point.

This work is organized as follows: We start by defining the extension of our classical model which enables us to drive the system out of equilibrium, and briefly recap the static critical properties of the model.
We then discuss the connection to the KZM and investigate our numerical results.
Using known values of the static critical exponents and the dynamic critical exponent, we rescale the data to obtain the universal non-equilibrium scaling function for the order parameter, susceptibility, and higher order cumulants up to fourth order.
Additionally, we provide our own estimate for the dynamic critical exponent extracted from non-equilibrium scaling behavior.
Finally, we introduce the finite system size as an additional scaling variable and demonstrate the crossover from the non-equilibrium finite-time scaling to the standard finite-size scaling behavior, which is also controlled by universal scaling functions.
The last section then gives a summary of our results, and highlights possible future directions.

\section{Model}
We simulate a $Z_2$ symmetric scalar field theory defined by the lattice Hamiltonian
\begin{equation}
    H = \sum_x a^d \left\{
    \frac{1}{2} \pi_x^2 + \frac{1}{2a^2} \sum_{y\sim x}\phi_x\phi_y + \left( \frac{m^2}{2} + \frac{d}{a^2}\right) \phi_x^2 + \frac{\lambda}{4!} \phi_x^4 + J(t) \phi_x
    \right\}.
\end{equation}
Here $\phi_x$ is a real scalar field variable at lattice site $x$, and $\pi_x$ its conjugate momentum. The lattice spacing is denoted by $a$ and will be set to unity in the following. If not stated otherwise, all quantities are given in corresponding lattice units. The sum $\sum_{y\sim x}$ runs over all nearest neighbors of $x$, and the sum $\sum_x$ runs over all lattice sites in the system. We will denote the number of lattice sites per dimension by $L$, the volume of the system is then given by $V = L^d$, where $d\in\{2,3\}$ counts the spatial dimensions.
Our model parameters will be set to $m^2 = -1$ and $\lambda = 1$.

The dynamics of the system is governed by Langevin-type equations of motion
\begin{equation}
    \partial_t \phi_x = \frac{\partial H}{\partial \pi_x}, \quad \partial_t \pi_x = -\frac{\partial H}{\partial \phi_x} - \gamma \pi_x + \sqrt{2\gamma T} \eta_x(t),
    \label{eq:eom1}
\end{equation}
where
\begin{equation}
    \frac{\partial H}{\partial \pi_x} = \pi_x, \quad \frac{\partial H}{\partial \phi_x} = -\sum_{y\sim x} (\phi_y - \phi_x) + \left(m^2 + \frac{\lambda}{6}\phi_x^2 \right) \phi_x + J(t).
    \label{eq:eom2}
\end{equation}
Here $\eta$ is a Gaussian white noise with zero mean and the variance given by $\langle \eta_x(t) \eta_y(t') \rangle = \delta_{xy} \delta(t-t')$. The parameter $\gamma$ is the dissipation constant, and $T$ is the temperature of the heat bath. We numerically solve the equations of motion using a leapfrog type scheme, as described in~\cite{Schweitzer:2021thesis}.

In the case of a vanishing external field, $J(t) = 0$, the system is symmetric under the transformation $\phi_x \to -\phi_x$. However, for temperatures $T<T_c$, the $Z_2$ symmetry is spontaneously broken and gets restored above the critical temperature $T_c$ via a second order phase transition. Since neither the order parameter nor the energy are conserved by the equations of motion (\ref{eq:eom1}) and (\ref{eq:eom2}), the system is said to be in the Model~A universality class in the Hohenberg-Halperin classification scheme~\cite{Hohenberg:1977ym}.

Observables of interest are the average order parameter or ``magnetization'' and the susceptibility, defined as
\begin{equation}
    \left\langle M \right\rangle  = \left\langle \frac{1}{V} \sum_x \phi_x \right\rangle ,
    \label{eq:magnetization}
\end{equation}
and
\begin{equation}
    \chi = \frac{V}{T} \left( \langle M^2 \rangle - \langle M \rangle^2 \right).
    \label{eq:susceptibility}
\end{equation}

\subsection{Static critical properties}

The static critical behavior of this model was studied extensively in the past.
As our setup is not optimized for the study of static critical properties, but rather for the study of non-equilibrium dynamics, we will not attempt to reproduce these results here and instead refer the reader to the literature.
We restate here the results relevant to this work for later reference.

For $d=2$ Onsager proposed an exact solution in 1944~\cite{Onsager:1943jn} from which the static critical exponents can be obtained analytically. For $d=3$ the static critical exponents have been determined numerically to high precision via the conformal bootstrap method~\cite{Kos:2016ysd,Komargodski:2016auf}. The static critical exponents are listed in \cref{tab:crit_exponents}.
The non-universal critical amplitudes of our model in both 2D and 3D were measured in~\cite{Schweitzer:2020noq}.
The results are listed in \cref{tab:crit_amplitudes} following the notation of~\cite{Pelissetto:2000ek}.

\begin{table}[ht] \centering
    \caption{\label{tab:crit_exponents}Static critical exponents of the 2D and 3D Ising model. In 2D the critical exponents are known exactly from Onsager's solution~\protect{\cite{Onsager:1943jn}}. In 3D the critical exponents were determined with high precision via the conformal bootstrap method~\protect{\cite{Kos:2016ysd,Komargodski:2016auf}}.
    }
    \begin{tabular}[t]{l S[table-format=2.3] S[table-format=1.8]}
        \toprule
                 & {$d=2$} & {$d=3$}               \\
        \midrule
        $\beta$  & 0.125   & 0.326419 \pm 0.000003 \\
        $\gamma$ & 1.75    & 1.237075 \pm 0.000010 \\
        $\delta$ & 15.0    & 4.78984 \pm 0.00001   \\
        $\nu$    & 1.0     & 0.629971 \pm 0.000004 \\
        $\omega$ & 2.0     & 0.82966 \pm 0.00009   \\
        \bottomrule
    \end{tabular}
\end{table}

\begin{table}[ht] \centering
    \caption{\label{tab:crit_amplitudes}Critical amplitudes (left) and correction amplitudes (right) of the 2D and 3D Ising model as determined in~\protect{\cite{Schweitzer:2020noq}}.
    }
    \begin{tabular}[t]{l S[table-format=1.6] S[table-format=1.7]}
        \toprule
              & {$d=2$}           & {$d=3$}             \\
        \midrule
        $T_c$ & 4.4629 \pm 0.0010 & 9.37074 \pm 0.00028 \\
        $B$   & 2.0203 \pm 0.0016 & 1.937 \pm 0.017     \\
        $B^c$ & 1.7425 \pm 0.0013 & 1.5291 \pm 0.0022   \\
        $C^c$ & 0.1222 \pm 0.0016 & 0.3173 \pm 0.0097   \\
        \bottomrule
    \end{tabular}
    \hspace{1em}
    \begin{tabular}[t]{l S[table-format=3.2] S[table-format=1.3]}
        \toprule
                & {$d=2$}      & {$d=3$}       \\
        \midrule
        $B_1^c$ & {---}        & 0.83 \pm 0.01 \\
        $B_2^c$ & 20.4 \pm 0.8 & {---}         \\
        $C_1^c$ & 302 \pm 48   & 2.2 \pm 0.4   \\
        \bottomrule
    \end{tabular}
\end{table}

When we move on to study non-equilibrium critical phenomena, we will normalize our input parameters and results using the non-universal amplitudes. If not explicitly stated otherwise, dimensionless scaling variables will be indicated by a bar. As such, we have the dimensionless symmetry breaking, magnetization and susceptibility constructed as
\begin{equation}
    \bar{J} \equiv J/J_0 = J(B^c/B)^\delta, \quad \bar{M} \equiv M/B, \quad \bar{\chi} \equiv \chi (\delta J_0/B).
\end{equation}
The dimensionless reduced temperature is defined as
\begin{equation}
    \tau \equiv \frac{T-T_c}{T_c}.
\end{equation}

We will also compare our results to the equilibrium equations of state of the order parameter and susceptibility given in~\cite{Schweitzer:2020noq}, as they are approached in the limit of an infinitely slow quench. These are given by
\begin{align}
    \langle M(\bar{J}) \rangle & = -\sgn(\bar{J}) B^c J_0^{1/\delta} |\bar{J}|^{1/\delta} \left( 1 + B_1^c |\bar{J}|^{\omega \nu_c} + B_2^c |\bar{J}| \right), \label{eq:equilibrium_M} \\
    \chi(\bar{J})              & = C^c J_0^{-\gamma_c} |\bar{J}|^{-\gamma_c} \left( 1 + C_1^c |\bar{J}|^{\omega \nu_c} \right),\label{eq:equilibrium_chi}
\end{align}
where $\nu_c\equiv \nu/\beta\delta$ and $\gamma_c \equiv \gamma/\beta\delta$. The critical amplitudes $B^c$, $C^c$ as well as the correction amplitudes $B_1^c$, $B_2^c$ and $C_1^c$ are listed in \cref{tab:crit_amplitudes}. Note that $J_0$ is chosen such that $B^c J_0^{1/\delta} = B$, and $C^c J_0^{-\gamma_c} = B/(\delta J_0)$.

\section{Non-equilibrium phase transitions}

The KZM describes the dynamical evolution of a system close to its critical point.
Since the systems' equilibrium relaxation time $\xi_t$ diverges close to criticality, any change of control parameters with a finite rate may lead to non-equilibrium behavior.
The non-equilibrium behavior then depends solely on universal properties of the given critical point,
as well as some macroscopic details of the trajectory in the phase diagram. In the spirit of~\cite{Chandran:2012cjk}, one can
distinguish three classes of such trajectories, which all start with an equilibrated system.

\begin{enumerate}
    \item trans-critical protocols (TCPs), which take the system across the critical point with a finite rate, smoothly interpolating the control parameters between to positions on opposite sides of the (continued) first-order transition,
    \item cis-critical protocols (CCPs), which merely let the system touch the critical point, staying in the same phase, and
    \item end-critical protocols (ECPs), where the system approaches the critical point asymptotically for~$t\to\infty$.
\end{enumerate}

In this study, we focus on linear TCPs where the control parameter is the external symmetry breaking field~$J(t)$ that will be changed at a constant rate~$r_J$ through the critical point, i.e.
\begin{equation}
    J(t) = -r_J t,
\end{equation}
where the time origin is chosen such that the critical point is reached at $t=0$.
The system is always prepared in a thermalized state at $J(t_i) = 0.005$ and $T=T_c$ and quenched to $J(t_f) = -J(t_i)$ at varying rates~$r_J>0$, keeping the temperature constant.

\begin{figure}[tb]
    \centering
    \includegraphics[width=.68\linewidth]{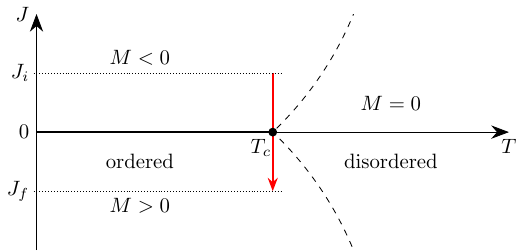}
    \label{fig:phase_diagram}
    \caption{Qualitative depiction of the phase diagram of the model and the trajectory of the quench protocol in the $T$-$J$ plane. The system is prepared in a negatively magnetized state at $J_i > 0$ and quenched to $J_f = -J_i$ at a constant rate $r_J$ through the critical point located at $J=0,\ T=T_c$ (black dot). The red arrow indicates the trajectory of the quench protocol. The thick black line represents the first-order transition line, while the dashed black lines show pseudo-critical lines between the ordered and disordered phases.}
\end{figure}

\subsection{Kibble-Zurek scaling relations}
As the system approaches the critical point at $J=0$, the correlation length $\xi$ and correlation time $\xi_t$ diverge according to the power laws
\begin{equation}
    \xi \sim |J|^{-\nu_c},
    \label{eq:xi}
\end{equation}
and
\begin{equation}
    \xi_t \sim |J|^{-\nu_c z}.
    \label{eq:xi_t}
\end{equation}
At some point, the relaxation process will be too slow for the system to follow the equilibrium state adiabatically, and the system will inevitably fall out of equilibrium.
The time at which the system falls out of equilibrium is defined to be the Kibble-Zurek time $t_\text{KZ}$.
The KZM assumes that the evolution of the systems stays adiabatic as long as the change in relaxation time $\dot{\xi}_t$ during the passage of a relaxation time $\xi_t$ is much smaller than the relaxation time $\xi_t$ itself~\cite{Chandran:2012cjk}, i.e.
$\dot{\xi}_t \xi_t <\!\!< \xi_t$ or $\dot{\xi_t} <\!\!< 1$.
One can then define the Kibble-Zurek time $t_\text{KZ}$ as the time when the change in relaxation time crosses this threshold,
\begin{equation}
    \dot{\xi}_t(t \equiv t_\text{KZ}) = 1.
    \label{eq:KZ_time}
\end{equation}
Exploiting the knowledge about the quench protocol and assuming that the system remains in equilibrium until $t_\text{KZ}$, we can write the relaxation time and its rate of change for times $t<0$ as
\begin{align}
    \xi_t(t)       & \sim (-r_J t)^{-\nu_c z},                                       \\
    \dot{\xi}_t(t) & \sim (-r_J t)^{-\nu_c z} (-t)^{-1} \nu_c  z \sim \xi_t(t)/(-t).
\end{align}
The rate of change of the relaxation time $\dot{\xi}_t$ becomes of order one when the time until the transition is of the same order as the equilibrium relaxation time, thus guaranteeing that the system falls out of equilibrium at a finite time
\begin{equation}
    t_\text{KZ} \sim - r_J^{-\nu_c z/(1+\nu_c z)}
    \label{eq:t_KZ}
\end{equation}
before the transition.
This also defines a characteristic scale $J_\text{KZ}$ for the external field at which the system falls out of equilibrium, i.e.
\begin{equation}
    J_\text{KZ} \equiv J(t_\text{KZ}) \sim r_J^{1/(1+\nu_c z)}.
    \label{eq:J_KZ}
\end{equation}

Now from this point onward, the system will no longer be able to follow the changing external parameter and the correlation length $\xi$ will be limited by the characteristic length scale $\xi_\text{KZ}$ set by the finite quench rate $r_J$ according to
\begin{equation}
    \xi_\text{KZ} \equiv \xi(t_\text{KZ}) \sim r_J^{-\nu_c/(1+\nu_c z)}.
    \label{eq:xi_KZ}
\end{equation}
The system stays out-of-equilibrium until the time $-t_\text{KZ}>0$ is reached, where the system can catch up again and will start to relax towards the new equilibrium state.

Note that while the authors of~\cite{Wu:2018twy} also investigated universal scaling functions of cumulants within the framework of Langevin dynamics of Model~A, they treat the dynamic critical exponent $z$ as an input to their model. They choose the Model~H value of $z=3$ and accordingly impose a critical slowing down of the relaxation time with the varying correlation length along the quench trajectory.
This is in contrast to our approach, where all model parameters are fixed and critical slowing down emerges naturally from the dynamics of the system, as the external symmetry breaking field drives the system through the critical point.
We therefore expect to find the dynamic critical exponent $z$ to have the well known Model~A value close to $z \approx 2$~\cite{Halperin:1972bwo}.
As a result, the universal scaling functions we compute will differ considerably from those found in~\cite{Wu:2018twy}, already on a qualitative level.

\section{Dynamic critical behavior}

\begin{figure}[tb]
    \centering
    \begin{subfigure}{.5\textwidth}
        \centering
        \includegraphics[width=.92\linewidth]{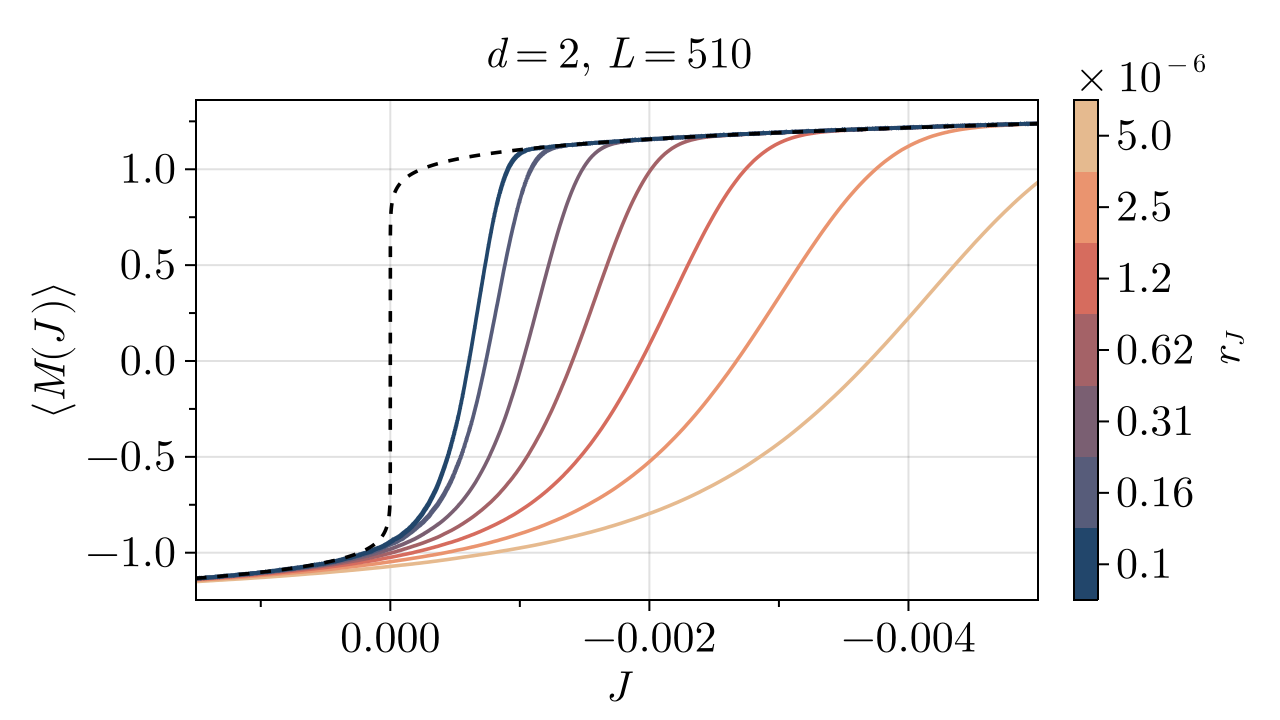}
        \label{fig:2D_M_unscaled}
    \end{subfigure}%
    \begin{subfigure}{.5\textwidth}
        \centering
        \includegraphics[width=.92\linewidth]{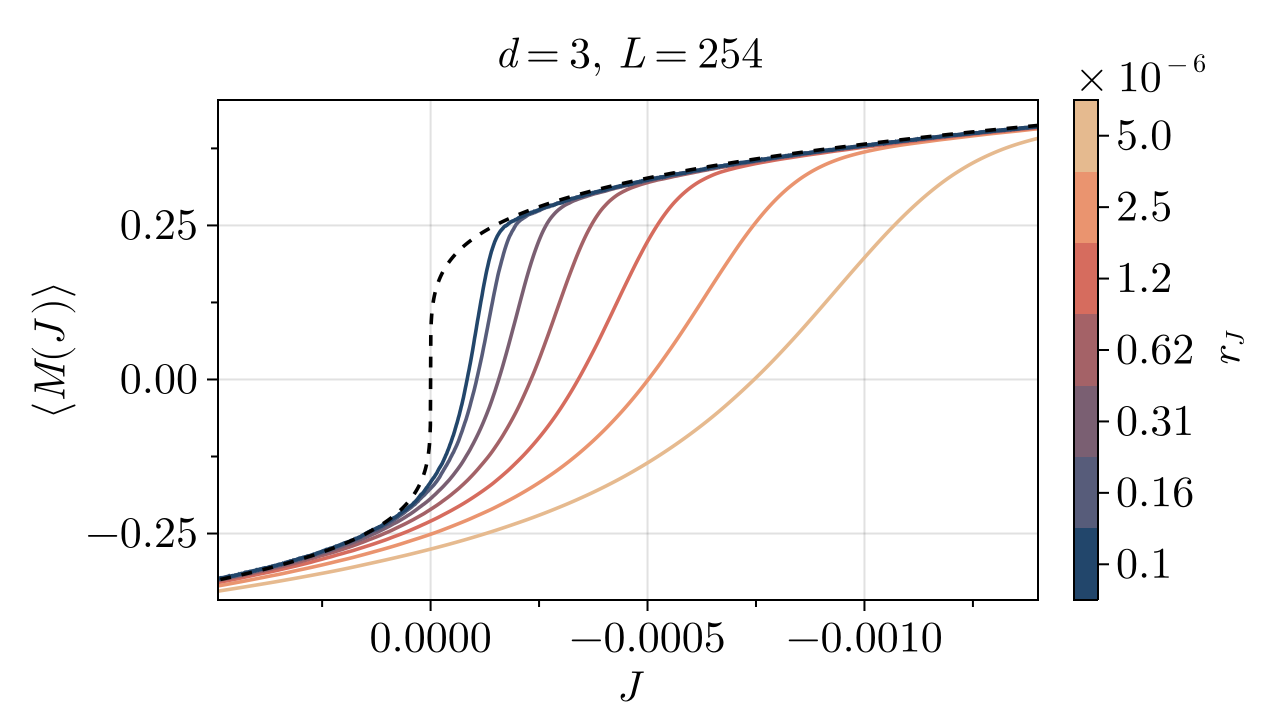}
        \label{fig:3D_M_unscaled}
    \end{subfigure}
    \begin{subfigure}{.5\textwidth}
        \centering
        \includegraphics[width=.92\linewidth]{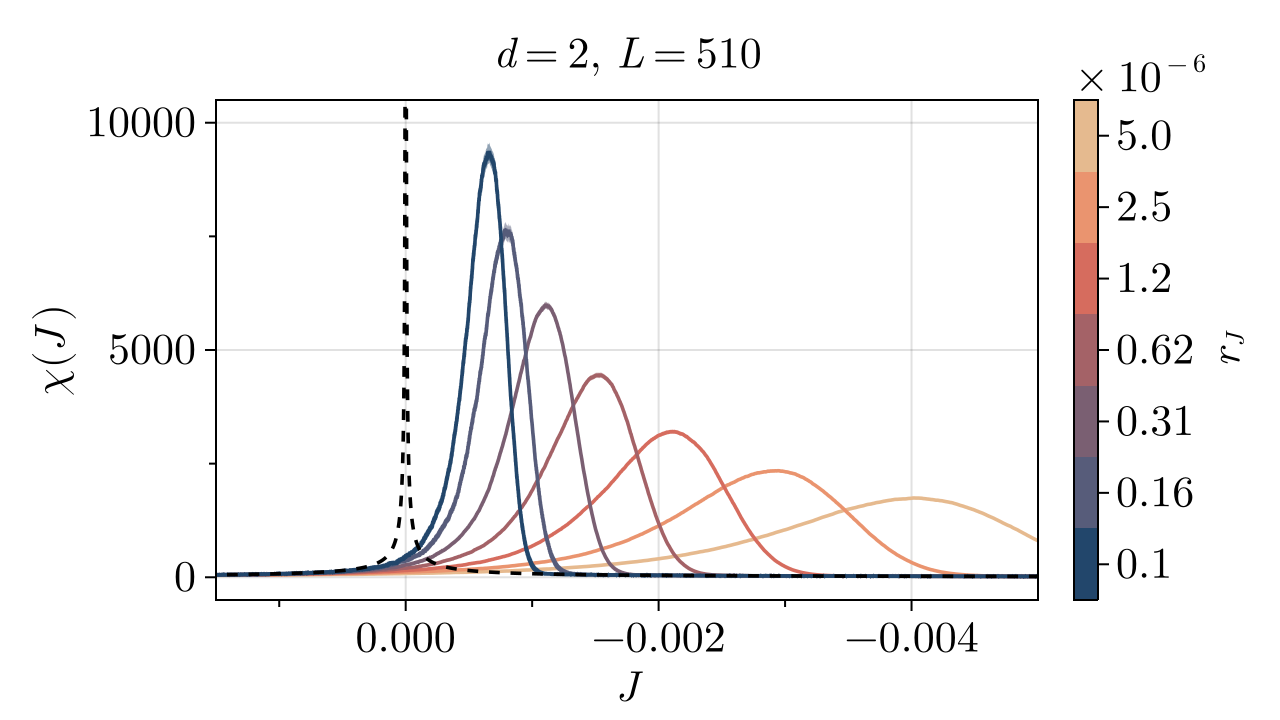}
        \label{fig:2D_chi_unscaled}
    \end{subfigure}%
    \begin{subfigure}{.5\textwidth}
        \centering
        \includegraphics[width=.92\linewidth]{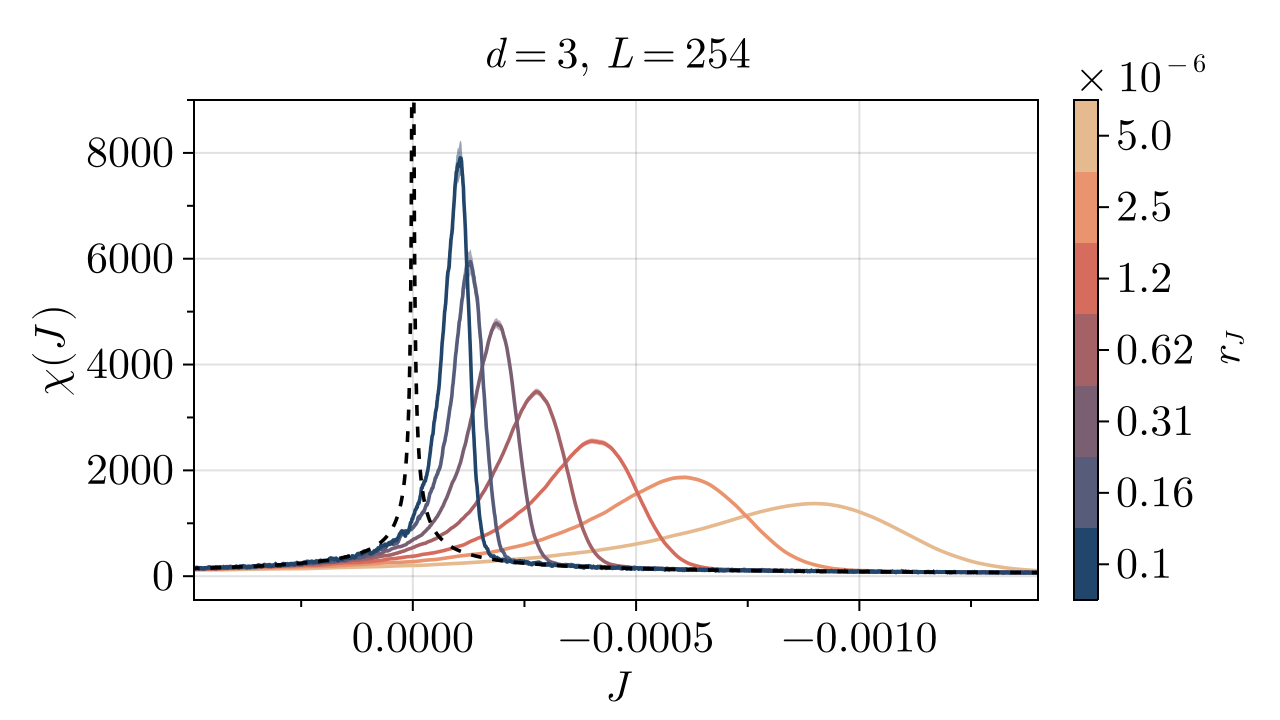}
        \label{fig:3D_chi_unscaled}
    \end{subfigure}
    \caption{Average magnetization $M$ (top) and susceptibility $\chi$ (bottom) as functions of $J$ at $T=T_c$ for different quench rates $r_J$.
        The dashed black lines correspond to the equilibrium equations of state (\ref{eq:equilibrium_M}) and (\ref{eq:equilibrium_chi}), which are approached in the limit $r_J\to 0$.
        Statistical uncertainties are presented as shaded areas around the curves, however, except for the smallest quench rates in case of the susceptibility, they are too small to be visible. Note that the horizontal axis is reversed to match the time direction of the quench protocol going from left to right.
    }
    \label{fig:unscaled_M_and_chi}
\end{figure}

We start the discussion of the dynamic critical behavior of the system by showing numerical results of the average magnetization $\langle M \rangle$ and susceptibility $\chi$ as functions of the external field $J$ for different quench rates $r_J$ in \cref{fig:unscaled_M_and_chi}. Different colors correspond to different quench rates.
We also include the equilibrium equations of state \cref{eq:equilibrium_M,eq:equilibrium_chi}, shown as dashed black lines, for comparison.
Averages denoted by $\langle \ldots \rangle$ are taken over the full ensemble of independent simulations, and statistical uncertainties are estimated using the bootstrap method.
These are presented as shaded areas around the curves, however, except for the smallest quench rates in case of the susceptibility, they are too small to be visible. For the results shown in \cref{sec:noneq_scaling_functions,sec:cumulants} around \num{10000} and \num{5000} independent simulations for each quench rate were performed in 2D and 3D, respectively. The finite-size scaling functions presented in \cref{sec:finite_size_scaling} were obtained by performing around \num{1000} additional simulations for each quench rate and multiple smaller system sizes.

The comparison of the magnetization for different quench rates with the equilibrium equation of state demonstrates a clear separation of two adiabatic regimes in the beginning and end of the quench, separated by a non-equilibrium regime around the transition point. The magnetization initially follows the equilibrium equation of state adiabatically, but starts to deviate from it before the critical point $J=0$ is reached, when the system falls out of equilibrium. The evolution of the magnetization then continues out-of-equilibrium along very different trajectories for different quench rates. The zero crossing of the order parameter is delayed to smaller values of $J$, and occurs further away from the critical point, with increasing quench rate.
At some point after the transition, the system starts relaxing towards the new equilibrium state, and the magnetization follows the equilibrium equation of state again. Qualitatively, the size of the non-equilibrium regime where the magnetization deviates from the equilibrium equation of state can be seen to increase with increasing quench rate as predicted by \cref{eq:J_KZ}.

Similar observations can be made for the susceptibility, while it is interesting to note that the susceptibility appears to no longer be symmetric around its maximum when the phase transition happens out-of-equilibrium. It appears that the transition from the initial adiabatic to the non-equilibrium regime happens more gradually than the return to equilibrium at the end of the quench.
The goal of the following analysis is to quantify these observations and capture the highly non-trivial out-of-equilibrium behavior of these observables in terms of universal scaling functions.

\subsection{Non-equilibrium scaling functions}
\label{sec:noneq_scaling_functions}

Scale invariance in the vicinity of a critical point allows us to write down the following general scaling ansatz for any observable $A$ with scaling dimension $\Delta_A$,
\begin{equation}
    A(J,r_J) = A_0 s^{\Delta_A} f_A(s^{1/\nu_c} \bar J, s^{z+1/\nu_c} \bar{r_J}).
    \label{eq:scaling_ansatz}
\end{equation}
Here $A_0$ is a non-universal amplitude, $s$ is an arbitrary parameter which rescales lengths as $l\to l/s$, and $f_A$ is a universal scaling function for which we adopt the normalization
\begin{equation}
    f_A(1,0)=f_A(0,1)=1.
    \label{eq:scaling_normalization}
\end{equation}
The dimensionless quench rate is defined as
$\bar{r_J} = r_J / {r_J}_0$
with the non-universal amplitude ${r_J}_0$ being determined such that the normalization condition $f_A(0,1)=1$ is satisfied.

By appropriately choosing the scaling parameter $s$, one can eliminate the dependence on one of the arguments and obtain a scaling function of only a single variable.
The first choice is to set $s=\bar{r_J}^{-\nu_c/(1+\nu_c z)}$, which eliminates the dependence on the second argument and yields
\begin{equation}
    A(J,r_J) = A_0 \bar{r_J}^{-\Delta_A\nu_c/(1+\nu_c z)} f_A(\bar x, 1),
    \label{eq:scaling_ansatz2}
\end{equation}
where we abbreviated the dimensionless scaling variable $\bar x = \bar{r_J}^{-1/(1+\nu_c z)} \bar J$.
Recalling the Kibble-Zurek scaling relation (\ref{eq:J_KZ}), we can also identify the scaling variable as $\bar x = J/J_\text{KZ}$.

A possible second choice would be to set $s=\bar{J}^{-\nu_c}$, which eliminates the dependence on the first argument, yielding
\begin{equation}
    A(J,r_J) = A_0 \bar{J}^{-\Delta_A\nu_c} f_A(1, \bar y), \quad\text{with } \bar y = \bar{J}^{-\nu_c z-1} \bar{r_J}.
    \label{eq:scaling_ansatz3}
\end{equation}
Note that both scaling functions $f_A(\bar x, 1)$ and $f_A(1, \bar y)$ are related via
\begin{equation}
    f_A(1, \bar y) =  \bar x^{\Delta_A\nu_c} f_A(\bar x, 1).
\end{equation}
Therefore, knowledge of one of them is sufficient to fully describe the non-equilibrium behavior of the system near the critical point for any quench rate.
In the following, we will focus on the first form of the scaling function $f_A(\bar{x}, 1)$ which we will refer to as just $f_A(\bar{x})$ for brevity.

The scaling dimensions of the magnetization and susceptibility are $-\beta/\nu$ and $\gamma/\nu$, respectively. According to \cref{eq:scaling_ansatz2}, the average magnetization and susceptibility then follow as
\begin{align}
    \langle M (J, r_J)\rangle & = - M_0 \bar{r_J}^{\,\beta/(\beta\delta+\nu z)} f_M(J/J_\text{KZ}), \label{eq:scaling_ansatz_M}       \\
    \chi(J,r_J)               & = \chi_0 \bar{r_J}^{-\gamma/(\beta\delta+\nu z)} f_\chi(J/J_\text{KZ}), \label{eq:scaling_ansatz_chi}
\end{align}
where $M_0 = B$ is the non-universal amplitude of the magnetization listed in \cref{tab:crit_amplitudes} and the non-universal amplitude of the susceptibility is $\chi_0 = C^c J_0^{-\gamma_c}= B/(\delta J_0)$.
The remaining non-universal amplitude ${r_J}_0$ is determined by the normalization condition $f_M(0) = 1$.
The numerical values we obtain are ${r_J}_0 = \num{3040}(410)$ in $d=2$ and ${r_J}_0 = 110(26)$ in $d=3$.

When all critical exponents and non-universal amplitudes are known, the scaling functions can be obtained using \cref{eq:scaling_ansatz_M,eq:scaling_ansatz_chi} by appropriately rescaling the data. This however also requires knowledge of the dynamic critical exponent $z$. Although the dynamic critical exponent of Model~A is not known exactly, many estimates exist in the literature. We will use the high-precision Monte Carlo estimates of $z=2.1667(5)$ for $d=2$ determined by Nightingale and Blöte~\cite{Nightingale:2000} and $z=2.0245(15)$ for $d=3$ as determined by Hasenbusch~\cite{Hasenbusch:2020}. An investigation of how the dynamic critical exponent $z$ can be extracted from the observed Kibble-Zurek scaling behavior in our data can be found in \ref{sec:z}.

\begin{figure}[tb]
    \centering
    \begin{subfigure}{.5\textwidth}
        \centering
        \includegraphics[width=.92\linewidth]{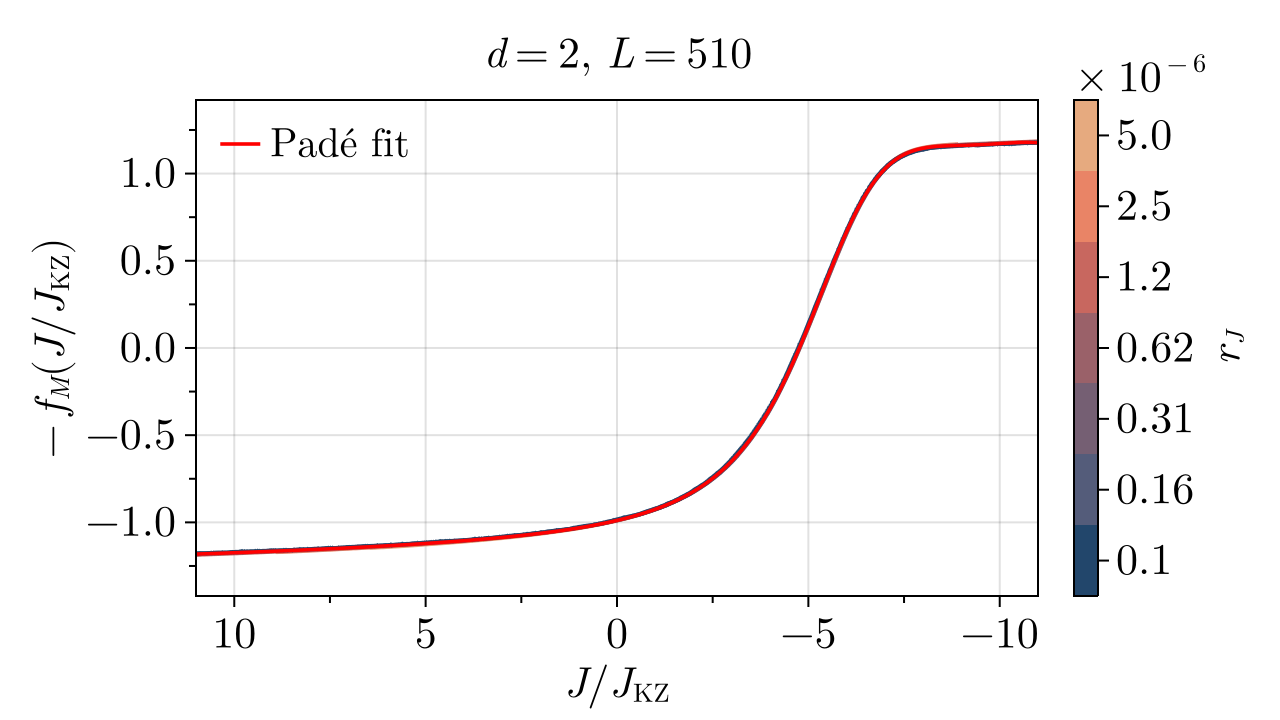}
        \label{fig:2D_N512_κ1_of_J_scaled}
    \end{subfigure}%
    \begin{subfigure}{.5\textwidth}
        \centering
        \includegraphics[width=.92\linewidth]{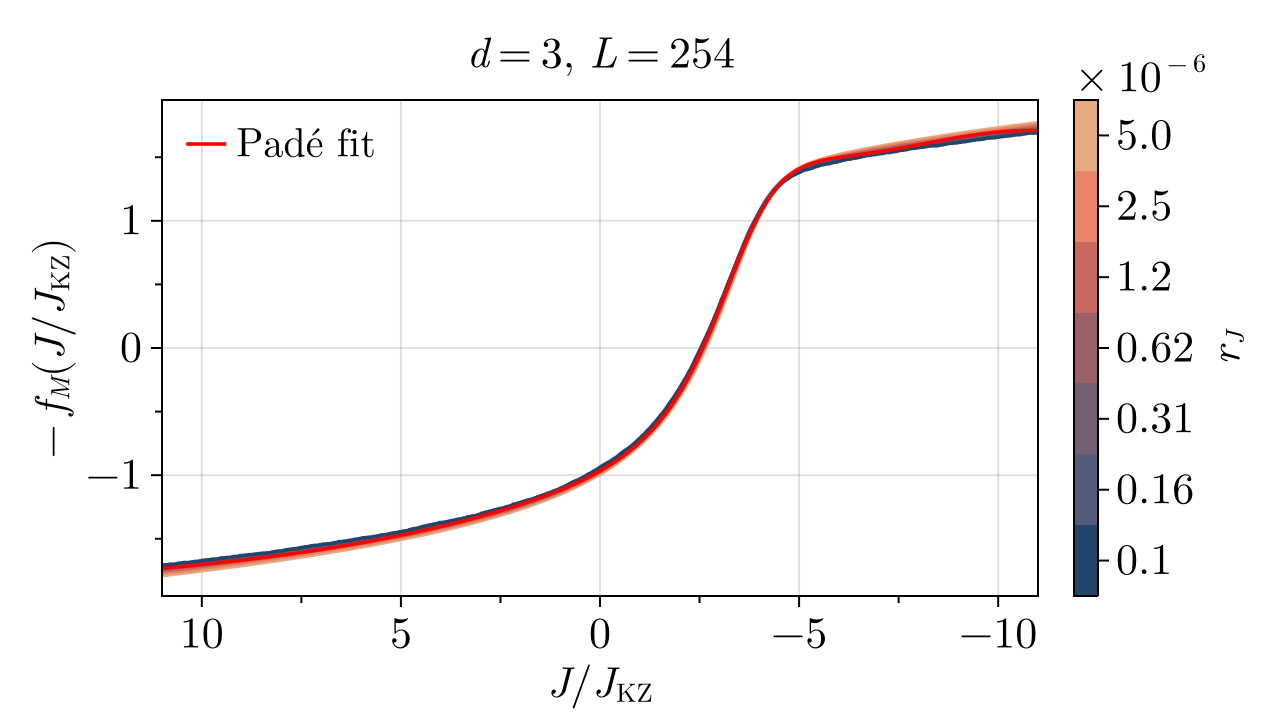}
        \label{fig:3D_N256_κ1_of_J_scaled}
    \end{subfigure}
    \begin{subfigure}{.5\textwidth}
        \centering
        \includegraphics[width=.92\linewidth]{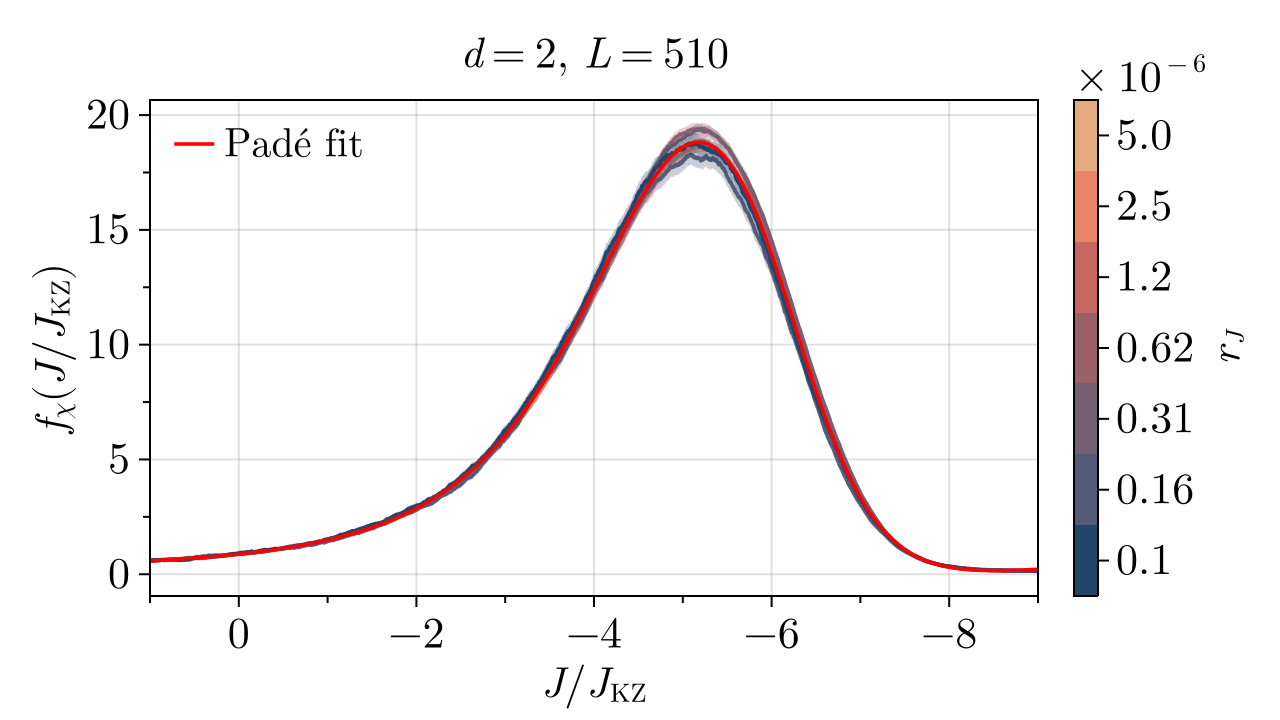}
        \label{fig:2D_N512_κ2_of_J_scaled}
    \end{subfigure}%
    \begin{subfigure}{.5\textwidth}
        \centering
        \includegraphics[width=.92\linewidth]{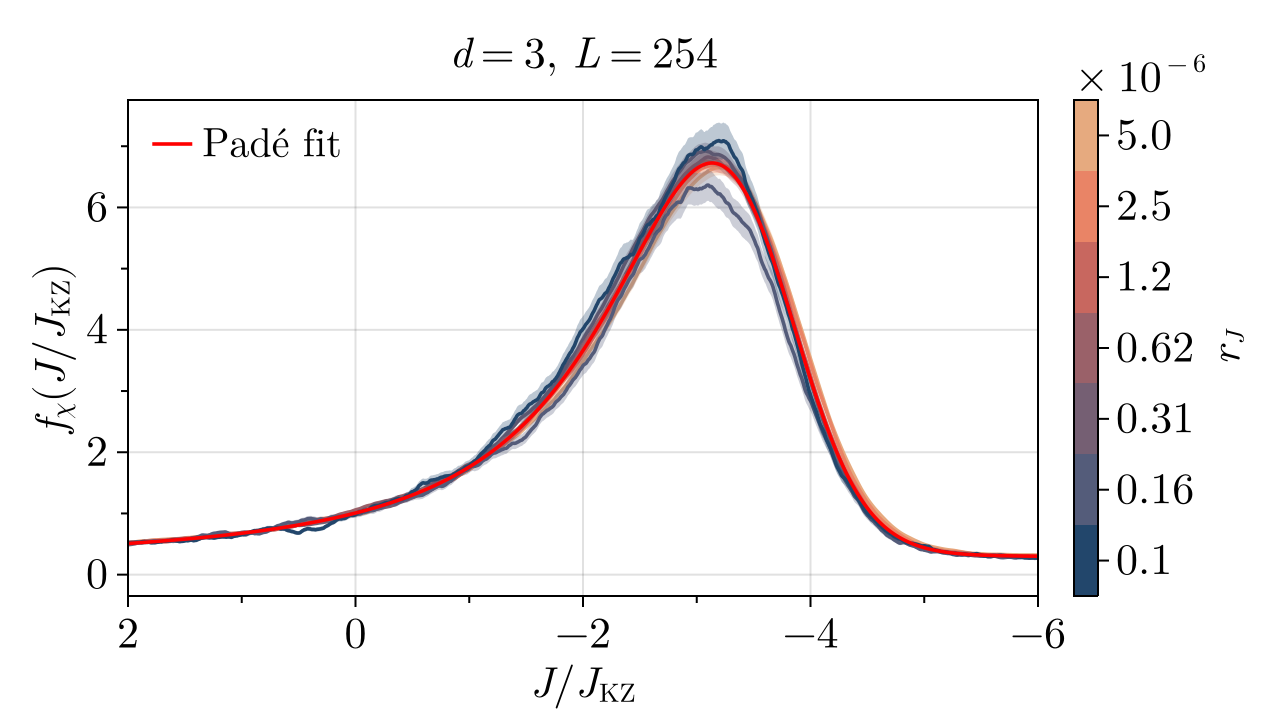}
        \label{fig:3D_N256_κ2_of_J_scaled}
    \end{subfigure}
    \caption{Order parameter $M$ (top) and susceptibility $\chi$ (bottom) as functions of the external field $J$ at $T=T_c$ for different quench rates $r_J$, scaled according to the dynamic scaling relations.
        All curves for different quench rates collapse rather well onto a single universal scaling function which was fit using a Padé approximant of order $[4/4]$ shown as a solid red line.}
    \label{fig:scaled_overview_M_and_chi}
\end{figure}

In \cref{fig:scaled_overview_M_and_chi} we show the rescaled magnetization and susceptibility as functions of the dimensionless scaling variable $\bar{x} = J/J_\text{KZ}$ for different quench rates.
A good collapse of the data for different quench rates onto a single curve is observed, indicating that the non-equilibrium evolution of the system near the critical point can indeed be described by a universal scaling function.

To provide closed form expressions for the scaling functions -- valid in the ranges presented in \cref{fig:scaled_overview_M_and_chi} -- we performed least squares fits of Padé approximants of order $[n/n]$
\begin{equation}
    R(x) = \frac{\sum_{j=0}^n a_j (x-x_0)^j}{1 + \sum_{k=1}^n b_k (x-x_0)^k}
\end{equation}
to the data.
The expansion point $x_0$ is matched to the region of interest which is around the zero crossing of the magnetization and maximum of the susceptibility, i.e.
$x_0=-5$ for $d=2$ and $x_0=-3$ for $d=3$.
We found that for both the magnetization and susceptibility, a Padé approximant of order $[4/4]$ is sufficient to describe the scaling function accurately in the intervals shown in \cref{fig:scaled_overview_M_and_chi}.
The resulting best estimates for the coefficients $a_j$ and $b_k$ are presented in \cref{tab:pade_2d,tab:pade_3d} for $d=2$ and $d=3$, respectively.

\subsection{Higher-order cumulants}
\label{sec:cumulants}

Higher-order cumulants of the order parameter such as the skewness and kurtosis are of special interest in the study of phase transitions as they are much more sensitive to the divergence of the correlation length~\cite{Stephanov:2008qz}.
In particular, cumulant ratios of the net-baryon number fluctuations have been proposed as observables to discover the QCD critical point using Heavy-Ion Collision experiments~\cite{Stephanov:1999zu}.
We will present universal scaling functions for the skewness and kurtosis of the order parameter in the following.

The skewness and kurtosis are defined as
\begin{equation}
    \kappa_3 = \left(\frac{V}{T}\right)^2 \left(\langle M^3 \rangle- 3 \langle M^2 \rangle \langle M \rangle + 2 \langle M \rangle^3 \right),
    \label{eq:kappa_3}
\end{equation}
and
\begin{equation}
    \kappa_4 = \left(\frac{V}{T}\right)^3 \left(\langle M^4 \rangle- 4 \langle M^3 \rangle \langle M \rangle - 3 \langle M^2 \rangle^2 + 12 \langle M^2 \rangle \langle M \rangle^2 - 6 \langle M \rangle^4 \right).
    \label{eq:kappa_4}
\end{equation}

A general formula for the scaling dimension of the $n$'th order cumulant $\kappa_n$ can be obtained from the equilibrium definition of the cumulant as
\begin{equation}
    \kappa_n^\text{eq} = \frac{\partial^{\,n-1} M}{\partial J^{n-1}} \sim |J|^{1/\delta - (n-1)}.
    \label{eq:kappa_n}
\end{equation}
Together with \cref{eq:xi} one can then obtain the scaling dimension of the $n$'th order cumulant to be
\begin{equation}
    \Delta_n = -\frac{\beta}{\nu} + \frac{n-1}{\nu_c}.
\end{equation}
For $n=1$ this reduces to the scaling dimension of the order parameter, $-\beta/\nu$. For $n=2$ we obtain the scaling dimension of the susceptibility, $\beta(\delta-1)/\nu=\gamma/\nu$, where the equality follows from Widom's identity~\cite{huang2008statistical}.
It is important to note that \cref{eq:kappa_n} is only valid in the equilibrium regime and determines the scaling dimension of the cumulants, while the cumulants as defined in \cref{eq:kappa_3,eq:kappa_4} are always valid observables in- and out-of equilibrium.

\begin{figure}[tb]
    \centering
    \begin{subfigure}{.5\textwidth}
        \centering
        \includegraphics[width=.92\linewidth]{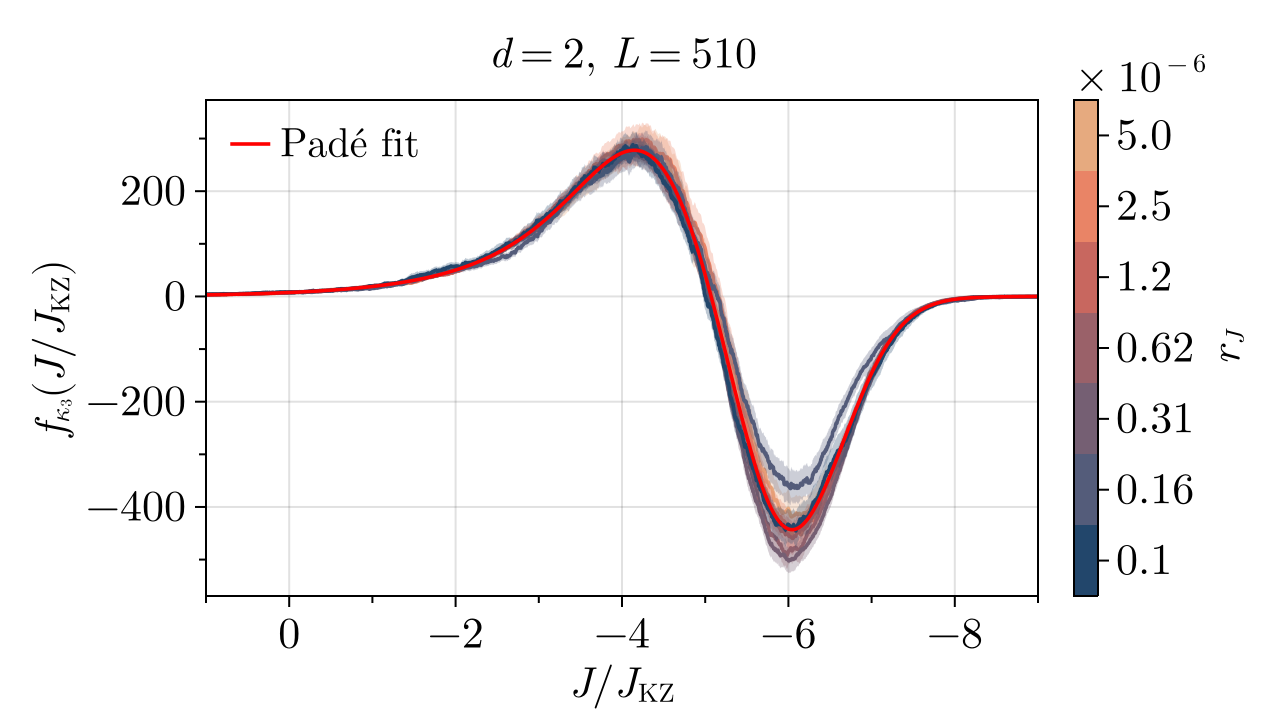}
        \label{fig:2D_N512_κ3_of_J_scaled}
    \end{subfigure}%
    \begin{subfigure}{.5\textwidth}
        \centering
        \includegraphics[width=.92\linewidth]{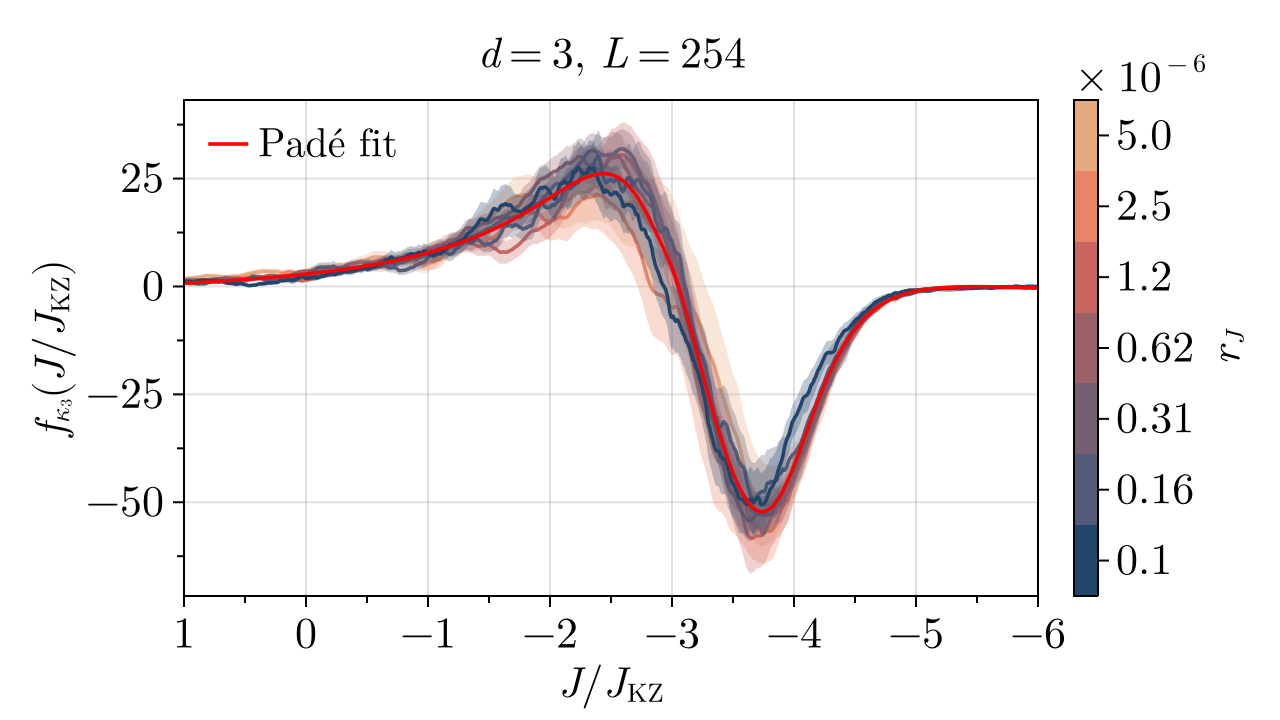}
        \label{fig:3D_N256_κ3_of_J_scaled}
    \end{subfigure}
    \begin{subfigure}{.5\textwidth}
        \centering
        \includegraphics[width=.92\linewidth]{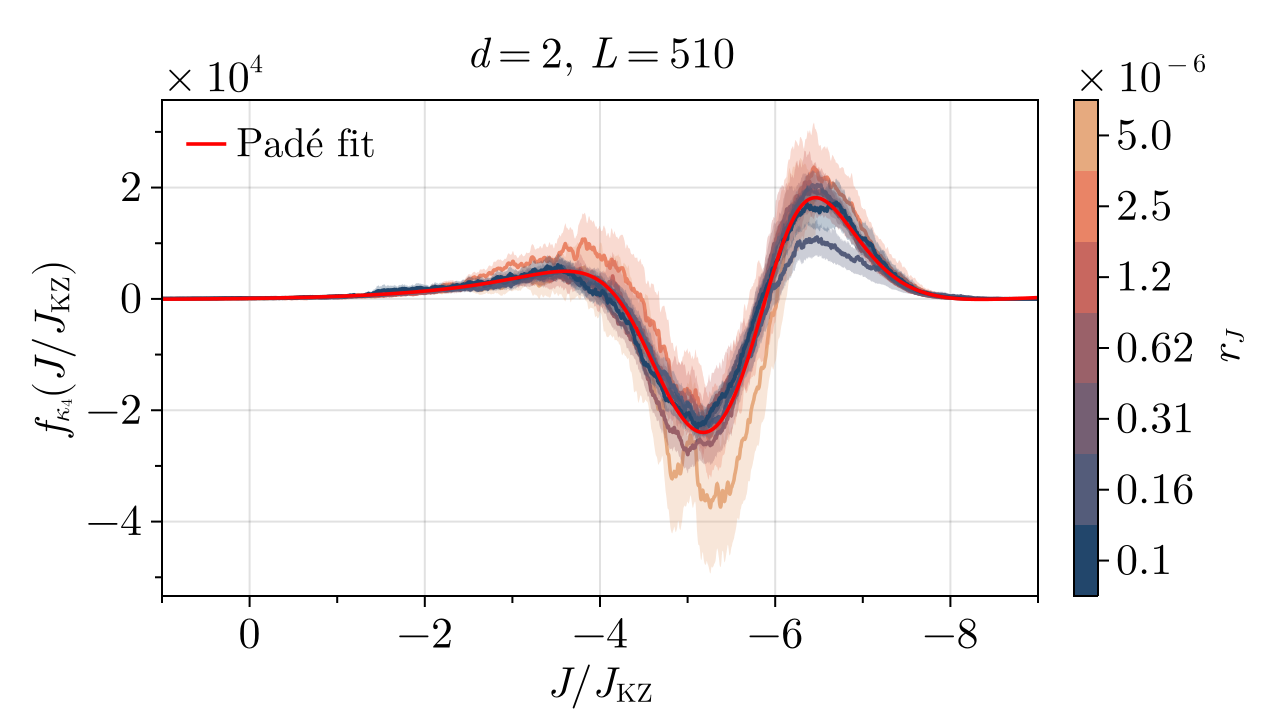}
        \label{fig:2D_N512_κ4_of_J_scaled}
    \end{subfigure}%
    \begin{subfigure}{.5\textwidth}
        \centering
        \includegraphics[width=.92\linewidth]{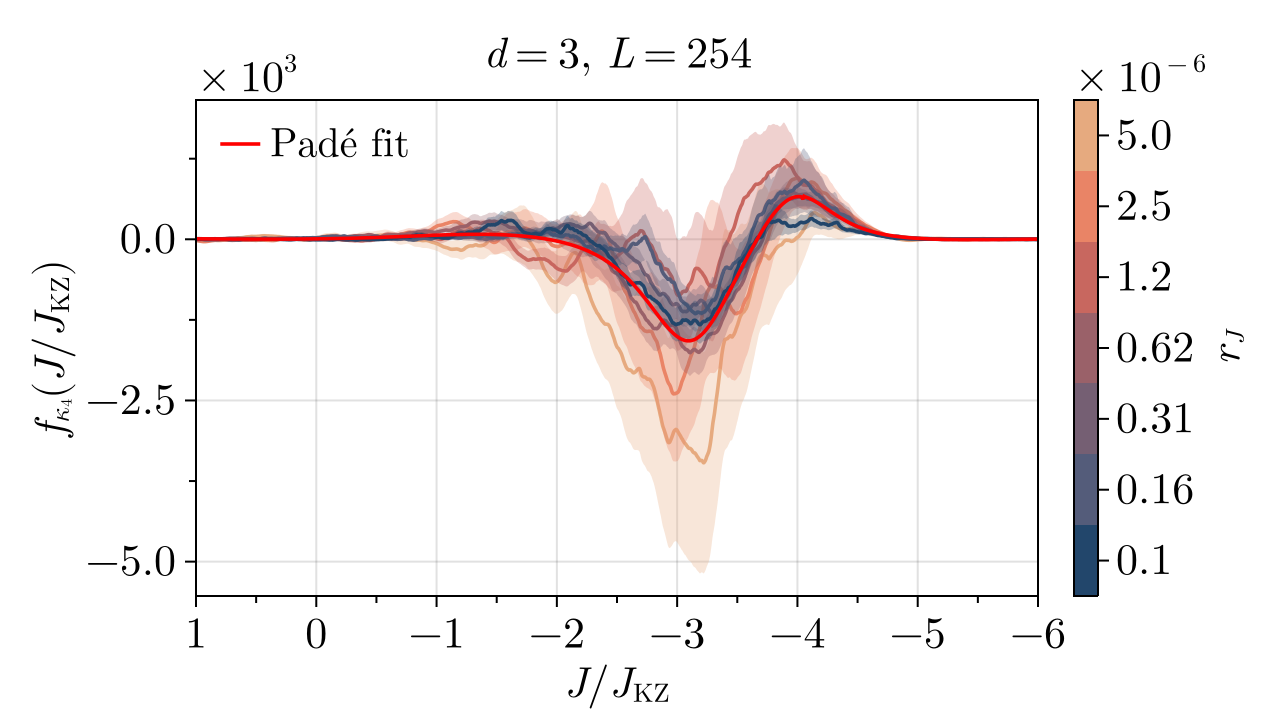}
        \label{fig:3D_N256_κ4_of_J_scaled}
    \end{subfigure}
    \caption{Skewness $\kappa_3$ (top) and kurtosis $\kappa_4$ (bottom) as functions of the external field $J$ at $T=T_c$ for different quench rates $r_J$, scaled according to the dynamic scaling relations. Universal scaling functions were fit using Padé approximants of order $[n/n]$, which are again shown as  solid red lines.}
    \label{fig:scaled_overview_k3_and_k4}
\end{figure}

We therefore expect the general scaling ansatz (\ref{eq:scaling_ansatz}) to also hold for the higher-order cumulants. Using the same values for the critical exponents as quoted previously, we rescale our data according to \cref{eq:scaling_ansatz2} and present the results for the scaling functions in \cref{fig:scaled_overview_k3_and_k4}.

Up to the statistical uncertainties in the data, we again find that the data for different quench rates collapses onto universal scaling functions.
The scaling functions for the skewness and kurtosis were fit by Padé approximants of orders $[6/6]$ and $[7/7]$ respectively with the best estimates for the coefficients being listed in \cref{tab:pade_2d,tab:pade_3d}.

\subsection{Finite-size scaling functions}
\label{sec:finite_size_scaling}

Up until now we have only considered the out-of-equilibrium behavior of systems in which the maximum correlation length reached during the quench is limited by the quench rate and not by the system size, i.e. ${\xi_\text{KZ} \sim r_J^{-\nu_c/(1+\nu_c z)} <\!\!< L}$. If this condition is not satisfied, the finite size of the system will have an effect on the evolution of the system.
If the finite size of the system is smaller than the correlation length at the Kibble-Zurek time, it will be frozen before the system falls out of equilibrium and the evolution stays adiabatic all the way through the transition. The scaling ansatz \cref{eq:scaling_ansatz2} should then be modified to include the finite size of the system as
\begin{equation}
    A(J, r_J,L^{-1}) = A_0 s^{\Delta_A} f_A(s^{1/\nu_c} \bar J, s^{z+1/\nu_c} \bar{r_J}, s \bar{L}^{-1}).
    \label{eq:scaling_ansatz_fss}
\end{equation}
Here, $\bar{L} = L/L_0$ is the dimensionless lattice size and $L_0$ is a constant reference length scale which can be determined from the equilibrium finite-size scaling behavior of the system and enforces the normalization condition $f_A(0,0,1)=1$.
Again, there are multiple choices for the scaling parameter $s$ which eliminate the dependence on one of the variables and yield a scaling function of two variables.
The choice we will focus on is $s=\bar{r_J}^{-\nu_c/(1+\nu_c z)}$, which eliminates the dependence on the second argument and yields
\begin{equation}
    A(J, r_J,L^{-1}) = A_0 \bar{r_J}^{-\Delta_A\nu_c/(1+\nu_c z)} f_A(\bar{x}, 1, \bar{z}), \quad \text{with } \bar{x} = \bar{r_J}^{-1/(1+\nu_c z)} \bar J,\ \bar{z} = \bar{r_J}^{-\nu_c/(1+\nu_c z)} \bar{L}^{-1}.
    \label{eq:scaling_ansatz_fss1}
\end{equation}
This universal scaling function $f_A(\bar{x}, 1, \bar{z})$ is the natural extension of the previously discussed ones to finite-size systems via the introduction of an additional scaling variable related to the system size.
In the limit of infinite system size, the new argument $\bar{z}$ will approach zero, and we recover the previously determined universal scaling function $f_A(J/J_\text{KZ}) \equiv f_A(J/J_\text{KZ}, 1, 0)$.
Knowledge of the two parameter scaling function is sufficient to fully describe the critical contribution to the non-equilibrium behavior of the system under consideration for any quench rate and system size.

\begin{figure}[tb]
    \centering
    \begin{subfigure}{.5\textwidth}
        \centering
        \includegraphics[width=.8\linewidth]{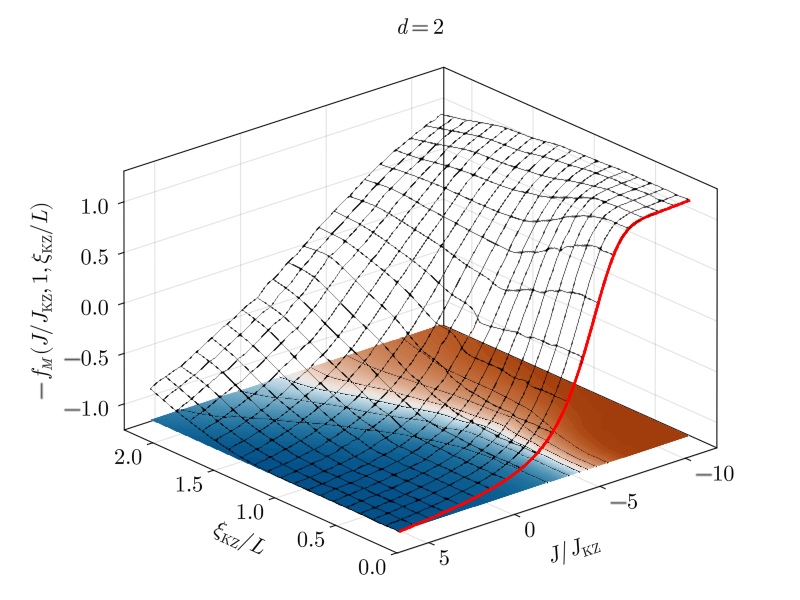}
        \label{fig:2D_FSS_M_with_data}
    \end{subfigure}%
    \begin{subfigure}{.5\textwidth}
        \centering
        \includegraphics[width=.8\linewidth]{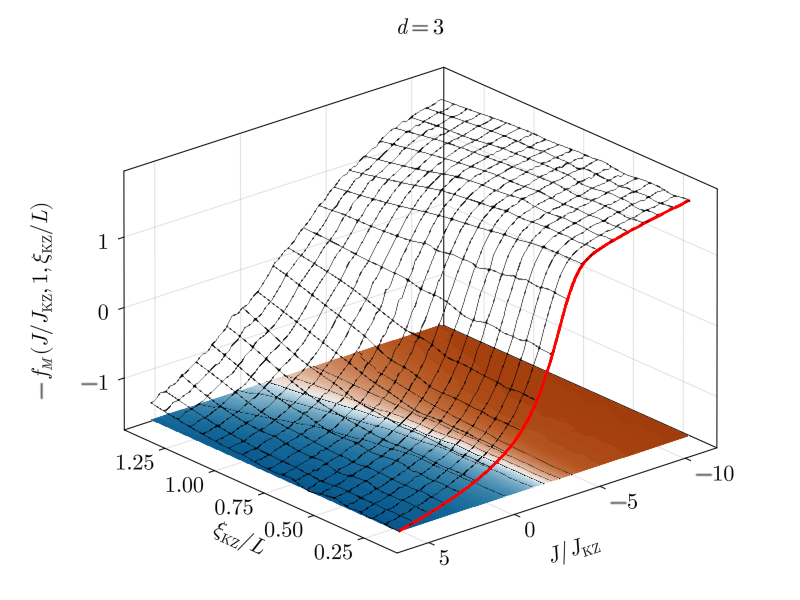}
        \label{fig:3D_FSS_M_with_data}
    \end{subfigure}
    \begin{subfigure}{.5\textwidth}
        \centering
        \includegraphics[width=.8\linewidth]{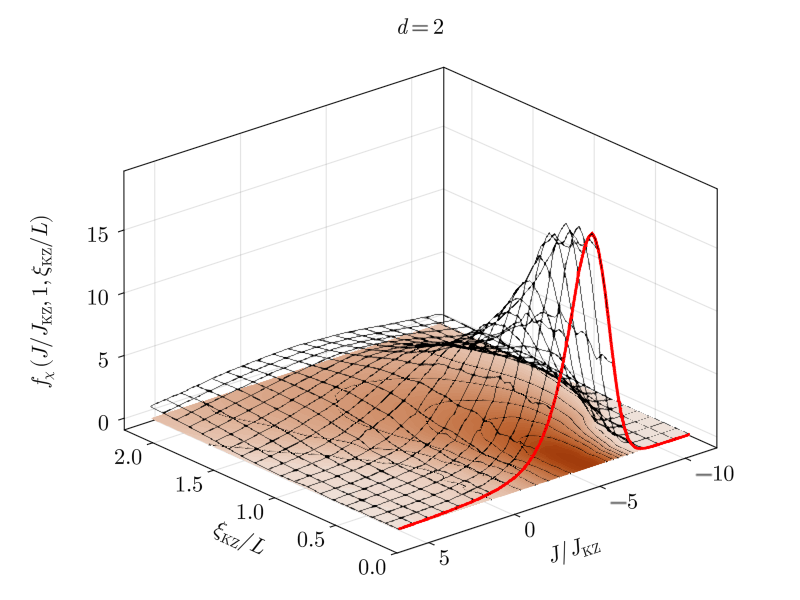}
        \label{fig:2D_FSS_chi_with_data}
    \end{subfigure}%
    \begin{subfigure}{.5\textwidth}
        \centering
        \includegraphics[width=.8\linewidth]{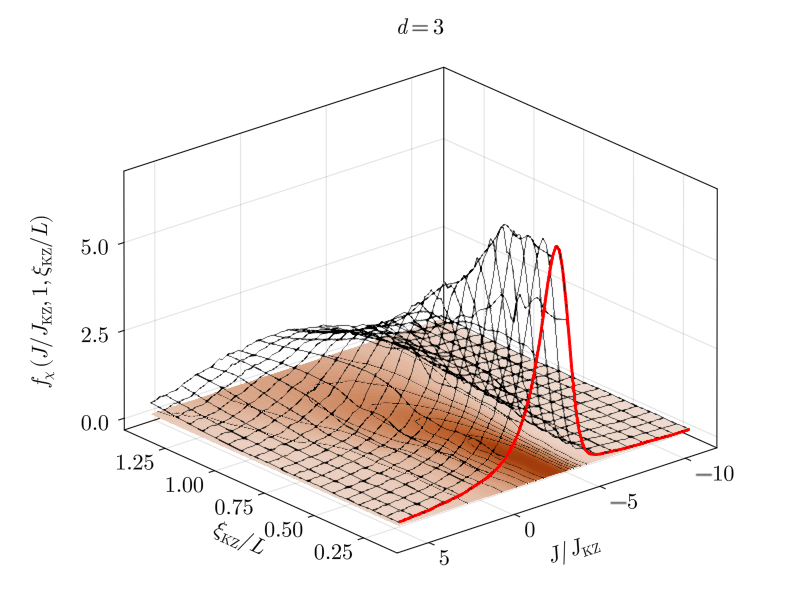}
        \label{fig:3D_FSS_chi_with_data}
    \end{subfigure}
    \begin{subfigure}{.5\textwidth}
        \centering
        \includegraphics[width=.8\linewidth]{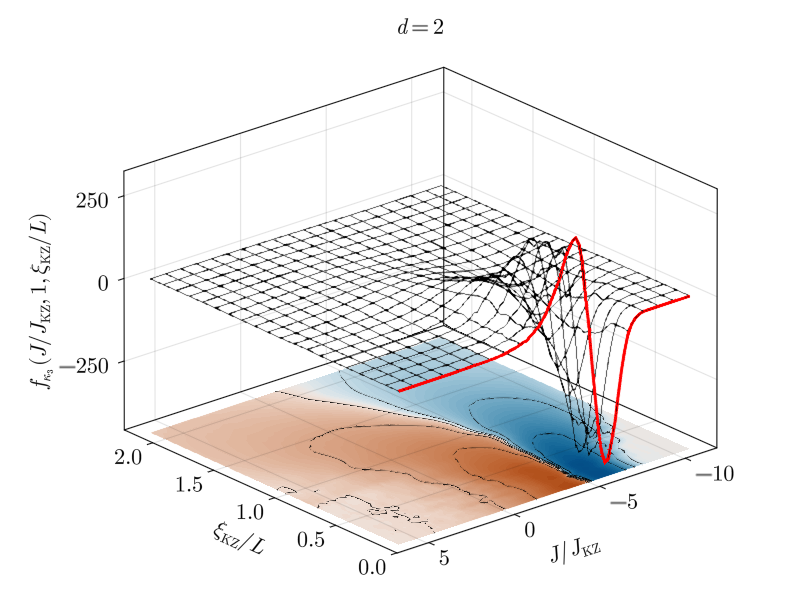}
        \label{fig:FSS_2D_κ3}
    \end{subfigure}%
    \begin{subfigure}{.5\textwidth}
        \centering
        \includegraphics[width=.8\linewidth]{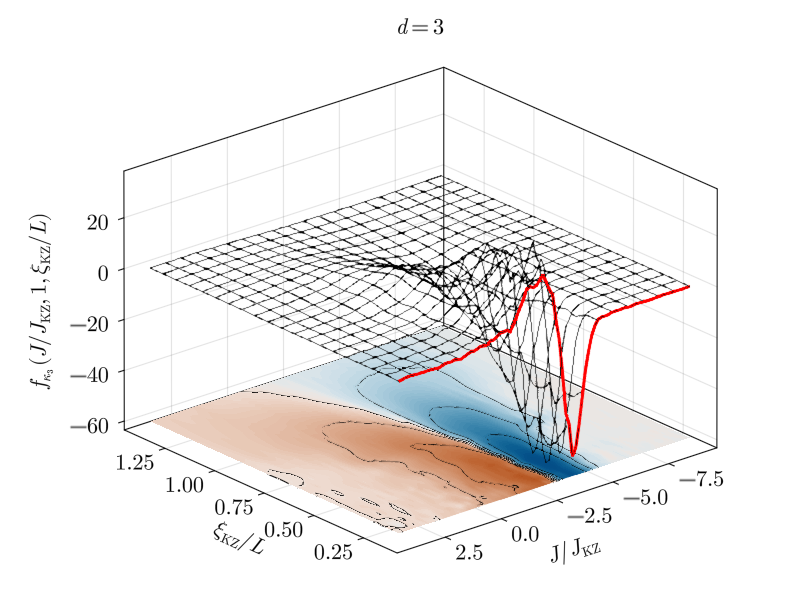}
        \label{fig:FSS_3D_κ3}
    \end{subfigure}
    \begin{subfigure}{.5\textwidth}
        \centering
        \includegraphics[width=.8\linewidth]{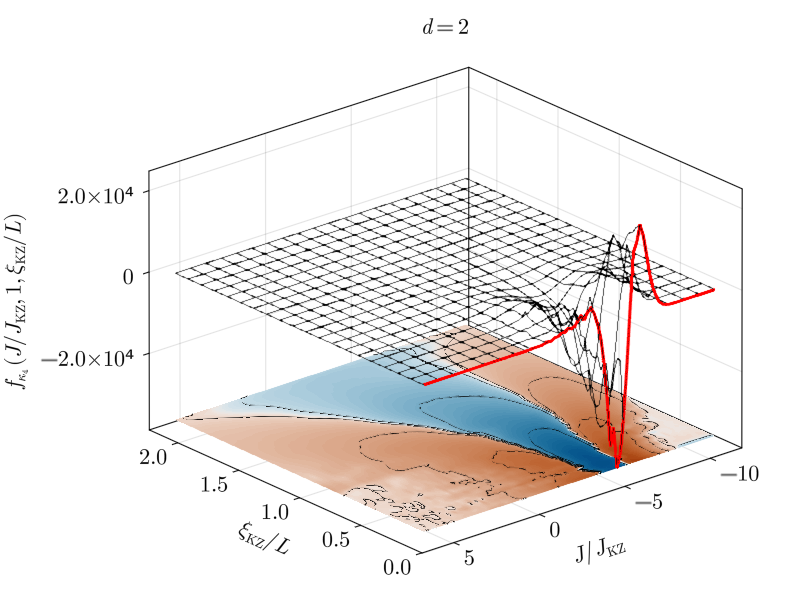}
        \label{fig:FSS_2D_κ4}
    \end{subfigure}%
    \begin{subfigure}{.5\textwidth}
        \centering
        \includegraphics[width=.8\linewidth]{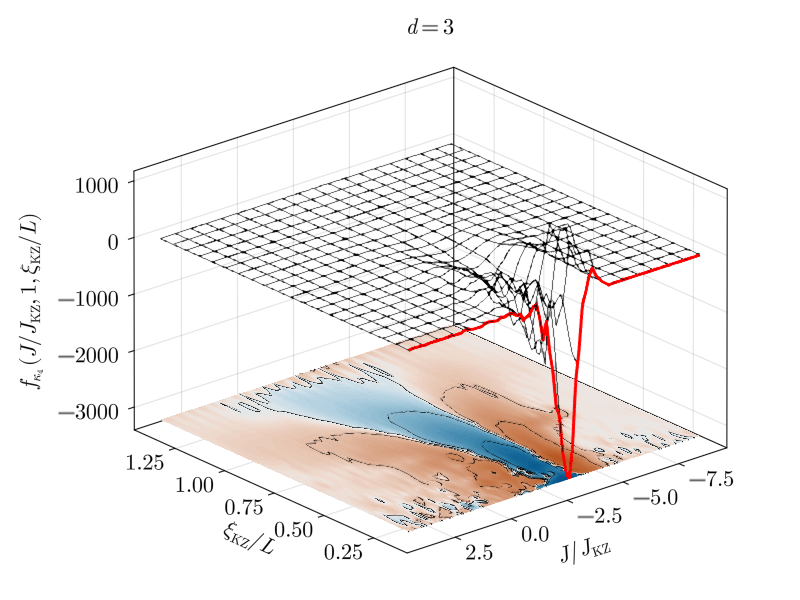}
        \label{fig:FSS_3D_κ4}
    \end{subfigure}
    \caption{Finite size scaling collapse of the average magnetization $\langle M \rangle$, susceptibility $\chi$, skewness $\kappa_3$ and kurtosis $\kappa_4$ (from top to bottom) at $T=T_c$. The wireframe shows a polyharmonic spline interpolation of the rescaled data revealing the universal scaling function.
        The red lines show the scaling function for the smallest available value of $\xi_\text{KZ}/L$.
        In the limit of infinite system size, these approach the universal scaling functions shown in \cref{fig:scaled_overview_M_and_chi} and \cref{fig:scaled_overview_k3_and_k4}.
    }
    \label{fig:FiniteSizeScaling}
\end{figure}

Again recalling the Kibble-Zurek scaling relations (\ref{eq:J_KZ}) and (\ref{eq:xi_KZ}), we can identify the scaling variables as $\bar{x} = J/J_\text{KZ}$ and $\bar{z} = \xi_\text{KZ}/L$.
Repeating the simulations we discussed previously for multiple different system sizes, we obtain slices of the scaling function $f_A(J/J_\text{KZ}, 1, \xi_\text{KZ}/L)$ for different values of $\xi_\text{KZ}/L$. A polyharmonic spline interpolation of the data is shown in \cref{fig:FiniteSizeScaling} as a wireframe, giving an overall view of the non-equilibrium finite-size scaling behavior of the order parameter and its cumulants.

\cref{fig:FiniteSizeScaling} shows that there is a strong dependence of the scaling function on the ratio of the Kibble-Zurek length scale to the system size. For small values of $\xi_\text{KZ}/L$, the scaling function approaches the previously determined universal scaling function of the approximately infinite system shown in \cref{fig:scaled_overview_M_and_chi,fig:scaled_overview_k3_and_k4} with the transition between the two oppositely ordered phases occurring at $J/J_\text{KZ} \approx -5$ and $J/J_\text{KZ} \approx -3$ for $d=2$ and $d=3$, respectively.
For values of $\xi_\text{KZ}/L \gtrsim 0.5$, corrections to the scaling behavior become visible. The transition starts to become less sharp, leading to a decrease in the susceptibility and the higher-order cumulants. Furthermore, the transition point shifts closer to $J/J_\text{KZ} = 0$ as the system size increases.
This is expected behavior as the evolution of the system becomes adiabatic in the limit of small system sizes and the system will not fall out of equilibrium during the quench protocol. We then recover the usual equilibrium finite-size scaling behavior of the system, where the transition still happens at $J=0$ and its steepness, again characterized by the maximum of the susceptibility, will scale with the system size.

\section{Conclusions}

We have investigated the non-equilibrium critical behavior in linear magnetic quenches across the critical point of a
relativistic scalar theory with a $Z_2$ order-parameter symmetry in 2+1D and 3+1D.
During the quench process, we observed the expected universal scaling behavior of the order parameter, transitioning from an initial equilibrium regime to a non-equilibrium regime governed by the Kibble-Zurek scaling laws and the dynamic critical exponent $z$.
The accuracy of our results proved sufficient to rescale the data and reveal the underlying universal non-equilibrium scaling functions governing the evolution of the order parameter and its higher-order cumulants.
Using fits to Padé approximants of the scaling functions, listed in \ref{sec:pade}, we are now able to predict the critical contribution of the time evolution of the order parameter and its cumulants in linear magnetic quenches across the critical point from an initial equilibrated state in a region around the critical point.

\ref{sec:z} provides a discussion on the extraction of the dynamic critical exponent $z$ from the observed Kibble-Zurek scaling behavior in our data.
Our results also show two possible ways in which Kibble-Zurek scaling can break down; in slow quenches, when the correlation length is limited by the finite system size, and in fast quenches, when the initial equilibrium relaxation time is too large for the system to follow the changing control parameter adiabatically.
By additionally considering the Kibble-Zurek length scale over the finite system size as a relevant scaling parameter, we were able to rescale the combined data for different lattice sizes and quench rates revealing the underlying universal non-equilibrium finite-size scaling functions as well.

With these tools, we are now well-equipped to explore more realistic non-equilibrium processes.
The most trivial extension would be to quench the system at a temperature close to, but not directly at the critical temperature, across the first order phase transition line and investigate how much of the time evolution is still given by the non-equilibrium universal function.
Similar studies have already been performed for 2D Potts models~\cite{Panagopoulos:2015uia,Pelissetto:2016tvy,Panagopoulos:2018hva}, $O(n)$ models~\cite{Scopa:2018rur} and the 2D Ising model~\cite{Zhong:2018,Fontana:2019zgc} and all found non-trivial off-equilibrium scaling behavior across first order phase transitions.
This study can be extended to more possible non-equilibrium processes, where both the heat-bath temperature and external field are changed continuously over time and closely pass the critical point.
Furthermore, employing a mapping between the variables of the QCD phase diagram and the 3D Ising model~\cite{Rehr:1973zz,Nonaka:2004pg,Parotto:2018pwx,Pradeep:2019ccv,Mroczek:2020rpm}, and considering more realistic transits of the QCD critical point~\cite{Berdnikov:1999ph,Akamatsu:2018vjr}, would greatly help in connecting the results of model studies to experimental observations in heavy-ion collisions.

Ultimately, to capture the full complexity of the non-equilibrium dynamics near the critical point, one would need to extend the model to the correct universality class of the QCD critical point which is conjectured to be Model~H in the classification of Hohenberg and Halperin~\cite{Hohenberg:1977ym, Son:2004iv}.
This is highly non-trivial from a numerical perspective.
However, recently simulations of stochastic fluid dynamics, displaying the correct Model~H behavior, have been successfully performed~\cite{Chattopadhyay:2024jlh}.
Re-investigating the non-equilibrium evolution of higher-order cumulants under these conditions would be especially interesting in the context of the search for the QCD critical point.

\section*{Acknowledgements}
We thank Jessica Fuchs, Frederic Klette and Johannes Roth for helpful discussions.
This work was supported by the Deutsche Forschungsgemeinschaft (DFG, German Research Foundation) through the CRC-TR 211 `Strong-interaction matter under extreme conditions' -- project number 315477589 -- TRR 211.

\appendix
\newpage
\section{Padé approximants of scaling functions}
\label{sec:pade}
\begin{table}[ht] \centering
    \caption{\label{tab:pade_2d} Best fit parameters for the Padé approximants of the scaling functions $f_M$, $f_\chi$, $f_{\kappa_3}$ and $f_{\kappa_4}$ around $J/J_\text{KZ}=5$ in two dimensions.}
    \begin{tabular}{l S[table-format=3.10] S[table-format=3.8] S[table-format=4.7] S[table-format=7.7]}
        \toprule
              & \multicolumn{4}{c}{$d=2$}                                                                      \\
              & {$f_M$}                   & {$f_\chi$}             & {$f_{\kappa_3}$}   & {$f_{\kappa_4}$}     \\
        \midrule
        $a_0$ & -0.12447  \pm 0.000071    & 18.6019   \pm 0.0023   & 40.96   \pm 0.2    & -22479  \pm 41       \\
        $a_1$ & 0.4943     \pm 0.00016    & 7.39      \pm 0.028    & 597.65  \pm 0.64   & 34020   \pm 650      \\
        $a_2$ & 0.14652    \pm 0.00045    & -0.064    \pm 0.015    & 334.5   \pm 6.1    & 41310   \pm 480      \\
        $a_3$ & 0.024136  \pm 0.00006     & -0.1266   \pm 0.0011   & 26.9    \pm 2.5    & -47600  \pm 1100     \\
        $a_4$ & 0.0014596 \pm 0.000007    & 0.02518   \pm 0.00045  & -11.13  \pm 0.8    & -19550  \pm 690      \\
        $a_5$ & {---}                     & {---}                  & -0.871  \pm 0.086  & 2810    \pm 360      \\
        $a_6$ & {---}                     & {---}                  & 0.165   \pm 0.027  & 801     \pm 28       \\
        $a_7$ & {---}                     & {---}                  & {---}              & -80     \pm 16       \\
        $b_1$ & 0.38877    \pm 0.00092    & 0.5229     \pm 0.0014  & 0.99    \pm 0.01   & -0.79     \pm 0.028  \\
        $b_2$ & 0.15192    \pm 0.00019    & 0.39231   \pm 0.00097  & 0.8162 \pm 0.0064  & -0.62     \pm 0.033  \\
        $b_3$ & 0.021115   \pm 0.000082   & 0.11242    \pm 0.00094 & 0.4124 \pm 0.0065  & -0.779     \pm 0.037 \\
        $b_4$ & 0.001073   \pm 0.0000043  & 0.05426   \pm 0.00036  & 0.2626  \pm 0.0047 & -0.673    \pm 0.04   \\
        $b_5$ & {---}                     & {---}                  & 0.1161 \pm 0.0033  & -0.919    \pm 0.038  \\
        $b_6$ & {---}                     & {---}                  & 0.0266  \pm 0.0011 & -0.474    \pm 0.033  \\
        $b_7$ & {---}                     & {---}                  & {---}              & -0.1332   \pm 0.0093 \\
        \bottomrule
    \end{tabular}
\end{table}

\begin{table}[ht] \centering
    \caption{\label{tab:pade_3d} Best fit parameters for the Padé approximants of the scaling functions $f_M$, $f_\chi$, $f_{\kappa_3}$ and $f_{\kappa_4}$ around $J/J_\text{KZ}=3$ in three dimensions.}
    \begin{tabular}[t]{l S[table-format=3.10] S[table-format=1.7] S[table-format=5.5] S[table-format=5.6]}
        \toprule
              & \multicolumn{4}{c}{$d=3$}                                                             \\
              & {$f_M$}                   & {$f_\chi$}          & {$f_{\kappa_3}$} & {$f_{\kappa_4}$} \\
        \midrule
        $a_0$ & -0.31211 \pm 0.00053      & 6.6366 \pm 0.0011   & 4.26 \pm 0.11    & -1507.7 \pm 4.8  \\
        $a_1$ & 0.6888 \pm 0.0011         & 5.853 \pm 0.033     & 73.2 \pm 4.2     & -2632 \pm 45     \\
        $a_2$ & 0.1149 \pm 0.0014         & 1.722 \pm 0.024     & 14 \pm 98        & 1124 \pm 83      \\
        $a_3$ & 0.04388 \pm 0.00021       & 0.1528 \pm 0.0035   & 2320 \pm 490     & 2852 \pm 48      \\
        $a_4$ & 0.00188 \pm 0.000024      & 0.04081 \pm 0.00065 & 1380 \pm 280     & 180 \pm 80       \\
        $a_5$ & {---}                     & {---}               & -27.2 \pm 8.5    & -519 \pm 34      \\
        $a_6$ & {---}                     & {---}               & -83 \pm 17       & -29.0 \pm 8.2    \\
        $a_7$ & {---}                     & {---}               & {---}            & 28.0  \pm 3.2    \\
        $b_1$ & 0.3021 \pm 0.0024         & 1.0758 \pm 0.0046   & 0.4 \pm 1.1      & 2.656 \pm 0.037  \\
        $b_2$ & 0.20464 \pm 0.00066       & 1.1115 \pm 0.0047   & 25.8 \pm 5.4     & 5.73 \pm 0.11    \\
        $b_3$ & 0.01605 \pm 0.00019       & 0.5162 \pm 0.0057   & 29.8 \pm 6.6     & 7.06 \pm 0.22    \\
        $b_4$ & 0.0010448 \pm 0.0000076   & 0.23 \pm 0.002      & 46.2 \pm 9.7     & 2.81 \pm 0.35    \\
        $b_5$ & {---}                     & {---}               & 46.5 \pm 9.5     & -0.97 \pm 0.28   \\
        $b_6$ & {---}                     & {---}               & 29.2 \pm 6.0     & -0.32 \pm 0.15   \\
        $b_7$ & {---}                     & {---}               & {---}            & 0.42 \pm 0.078   \\
        \bottomrule
    \end{tabular}
    \vspace{-2ex}
\end{table}
\newpage

\section{Extracting the dynamic critical exponent}
\label{sec:z}

In theory, the dynamic critical exponent $z$ can be determined by observing the Kibble-Zurek scaling of any characteristic feature of any observable in the non-equilibrium regime. For example, the value of the external field at which the magnetization crosses zero is expected to scale as
\begin{equation}
    J_{M=0} \sim r_J^{1/(1+\nu_c z)},
    \label{eq:J_M0}
\end{equation}
according to \cref{eq:J_KZ}.
Equivalently, one could also consider the time at which the magnetization crosses zero, which exhibits similar scaling behavior according to \cref{eq:t_KZ}.
Using the susceptibility instead, the location of the maximum of the susceptibility, for example, follows exactly the same scaling relations while according to \cref{eq:scaling_ansatz_chi} its amplitude is expected to scale as
\begin{equation}
    \chi_{\text{max}} \sim r_J^{-\gamma/(\beta\delta+\nu z)}.
    \label{eq:chi_max}
\end{equation}
Similar scaling relations also apply to the higher-order cumulants.
From power law fits to the data, one can then extract the dynamic critical exponent $z$ if all relevant static critical exponents are known.

\begin{figure}[tb]
    \centering
    \begin{subfigure}{.5\textwidth}
        \centering
        \includegraphics[width=.92\linewidth]{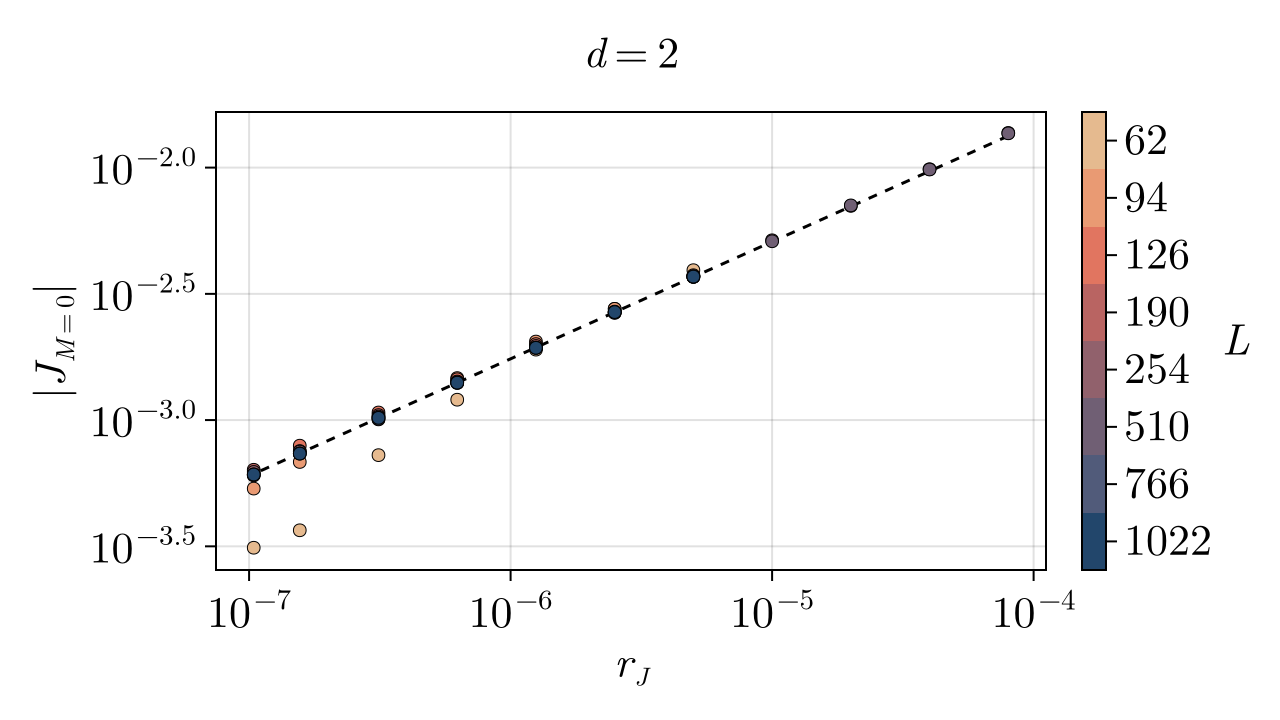}
        \label{fig:2D_fit_J}
    \end{subfigure}%
    \begin{subfigure}{.5\textwidth}
        \centering
        \includegraphics[width=.92\linewidth]{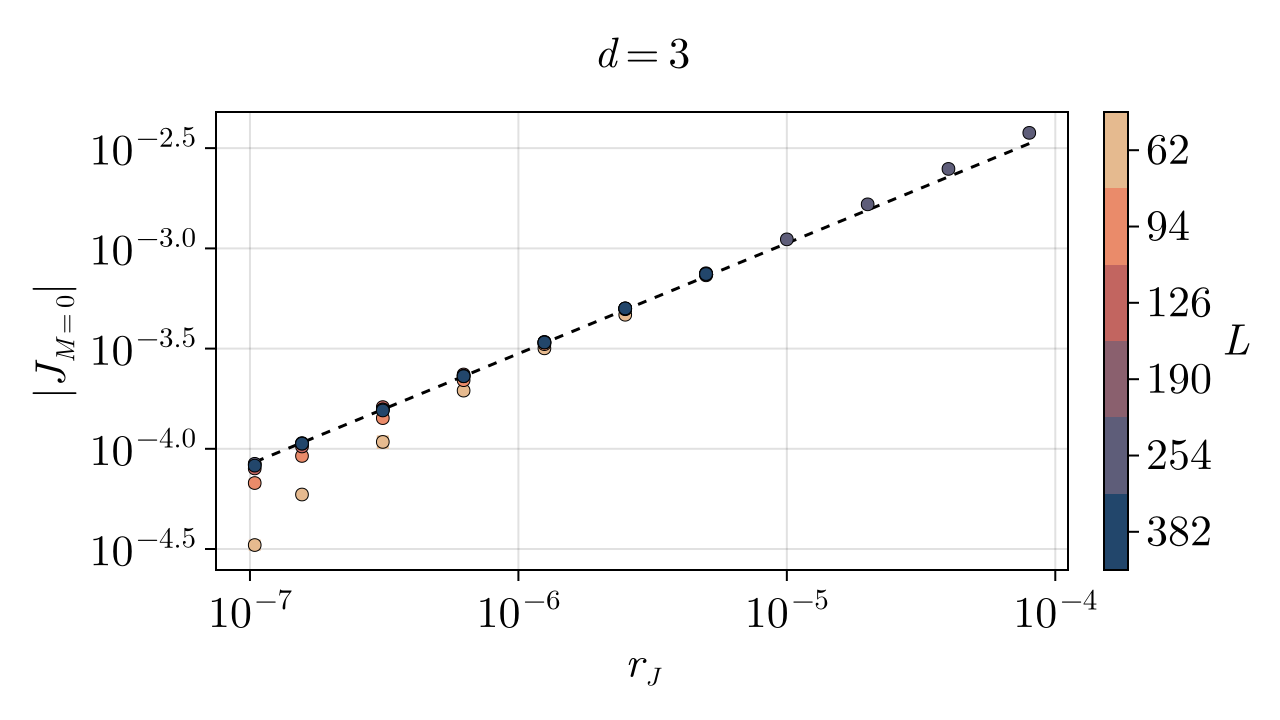}
        \label{fig:3D_fit_J}
    \end{subfigure}
    \caption{Zero crossing of the magnetization $J_{M=0}$ at $T=T_c$ as a function of the quench rate $r_J$ on a double logarithmic scale. The dashed line shows the expected scaling behavior with the exponent $1/(1+\nu_c z)$ according to \cref{eq:J_KZ}. Differently colored data points correspond to different lattice sizes. The error bars are much smaller than the data points and not visible in the plot.}
    \label{fig:fit_J_KZ}
\end{figure}

In practice however, this is a non-trivial task, as the data is usually noisy and any statistical uncertainties in the data will propagate into the fit results. This makes the zero crossing of the order parameter the best suited feature to determine the dynamic critical exponent, as it fluctuates the least and is easily identifiable in the data.
We present in \cref{fig:fit_J_KZ} the value of the external field at which the magnetization crosses zero as a function of the quench rate $r_J$ for different lattice sizes $L$. The data is plotted on a double logarithmic scale, and the expected scaling behavior according to \cref{eq:J_KZ} is shown as a dashed line.
The $d=2$ data clearly follows the expected power law behavior, while the $d=3$ data allows us to identify two ways in which the Kibble-Zurek scaling can break down.
First, if the quench rate is very small, the correlation length will be limited by the finite system size rather than the Kibble-Zurek length scale (\ref{eq:xi_KZ}) set by the quench rate. This is visible in both the $d=2$ and $d=3$ data for the smallest lattice sizes.
Second, if the quench rate is very large, Kibble-Zurek scaling is also expected to break down as the equilibrium relaxation time of the system will be too large for the system to follow the changing control parameter adiabatically during any part of the quench-protocol~\cite{Zeng:2022hut} and the system falls out of equilibrium before entering the critical region.

In both extremes, the observables become completely independent of the quench rate. However, before that happens, sub-leading and regular corrections to the expected Kibble-Zurek scaling laws can already influence the results.
Therefore, in any finite system the range of quench rates in which one can expect to observe Kibble-Zurek scaling will also be finite. Furthermore, the size of this dynamic scaling region is a non-universal quantity and therefore depends on the specific system under consideration.
It is known that leading order finite size corrections can be reduced by tuning the coupling constant $\lambda$ to an optimal value~\cite{Hasenbusch:1999}. We briefly investigated this for our model, but found no significant improvement in the scaling behavior of the data.

\begin{figure}[tb]
    \centering
    \begin{subfigure}{.5\textwidth}
        \centering
        \includegraphics[width=.92\linewidth]{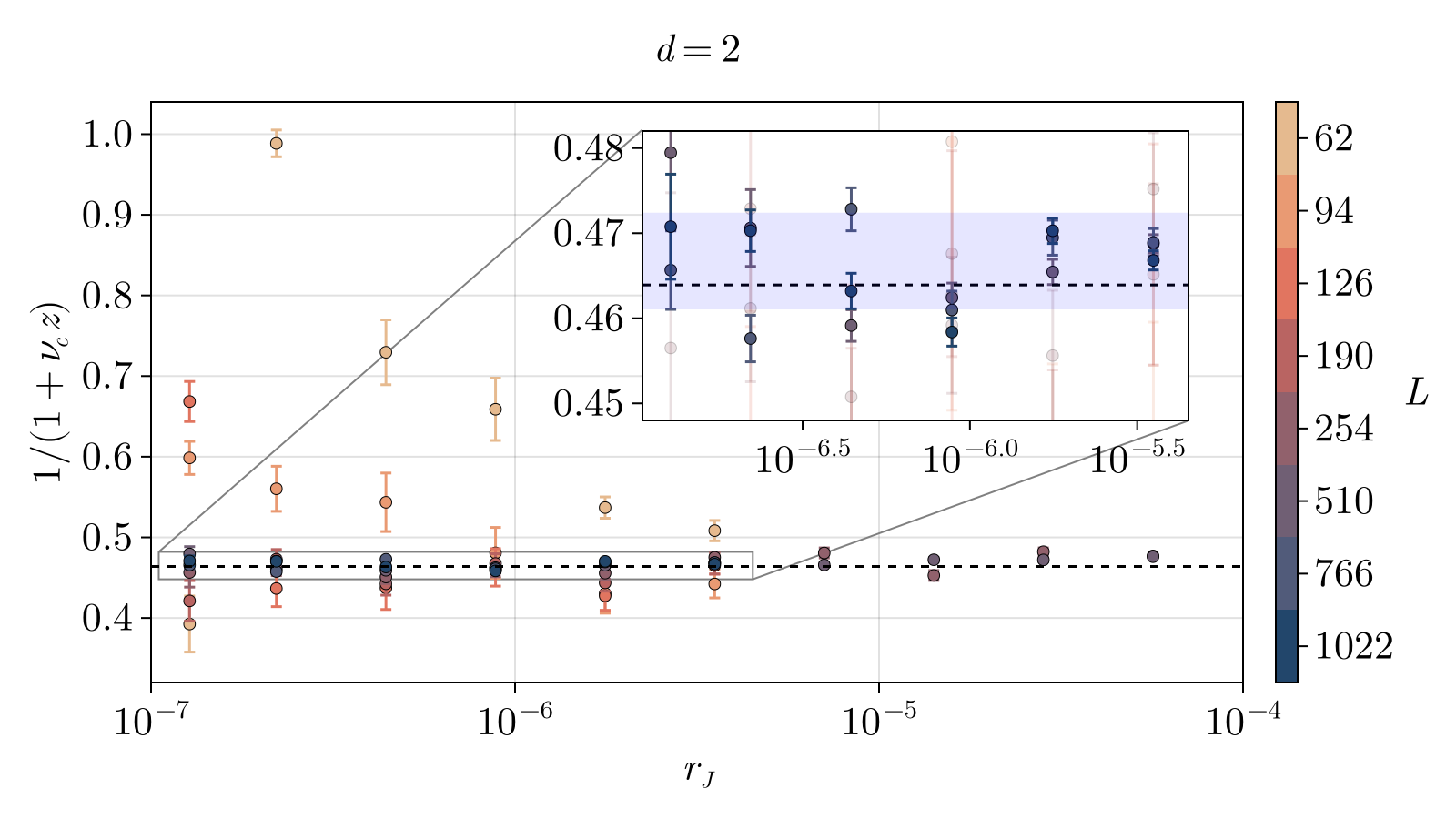}
        \label{fig:z_linregr_2D}
    \end{subfigure}%
    \begin{subfigure}{.5\textwidth}
        \centering
        \includegraphics[width=.92\linewidth]{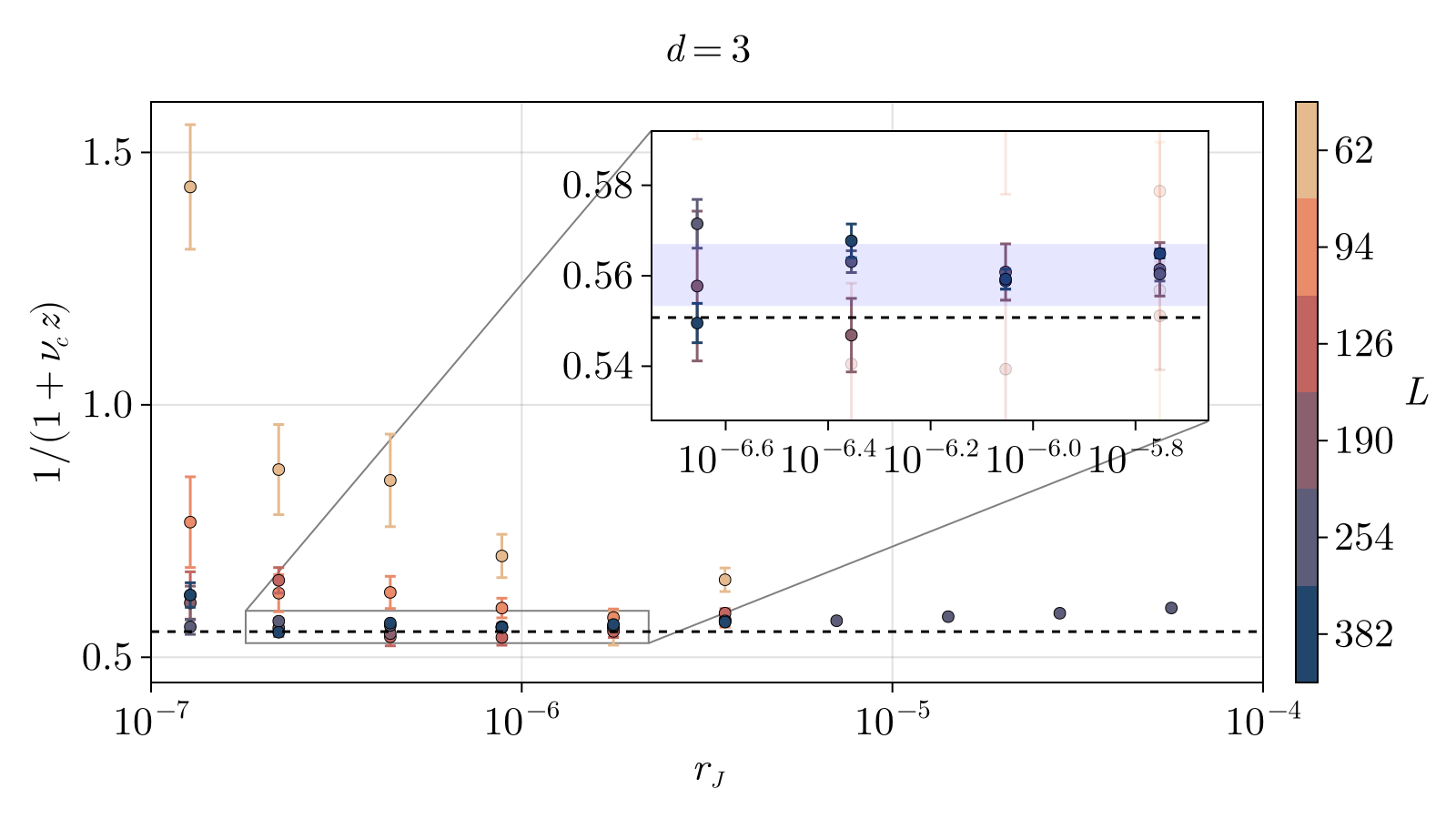}
        \label{fig:z_linregr_3D}
    \end{subfigure}
    \caption{Local slope estimates of the double logarithmic scaled data shown in \cref{fig:fit_J_KZ} corresponding to the scaling exponent $1/(1+\nu_c z)$ for the two-dimensional (left) and three-dimensional (right) case. The inset shows the plateau region of the data which was used for estimating the scaling exponent. Only the three largest lattice sizes ($L \in [510, 766, 1022]$ for $d=2$ and $L \in [190, 254, 382]$ for $d=3$) were used for the estimate which together with its standard deviation is shown as the blue band in the inset.
        Data points which were not used for the estimate are transparent in the inset.
        The dashed black line shows the expected scaling exponent derived from literature values for the static and dynamic critical exponents. The uncertainty of the derived literature value is smaller than the line width and not shown.
    }
    \label{fig:z_linregr}
\end{figure}

To better visualize the dynamic scaling region, we show in \cref{fig:z_linregr} the local slope estimates of the data shown in \cref{fig:fit_J_KZ} for the two-dimensional and three-dimensional case. These estimates are obtained by performing a linear regression on neighboring data points in the double logarithmic scaled data. If the data follows the expected power law behavior, the slope of the data should be constant and correspond to the scaling exponent $1/(1+\nu_c z)$.
The dashed black line shows the expected value of the scaling exponent according to the previously cited literature values for the static and dynamic critical exponents.
The inset in \cref{fig:z_linregr} shows the region were the slope estimates are approximately constant, which we identify as the dynamic scaling region.
We use our data from the three largest lattice sizes to estimate the dynamic scaling exponent for which we obtain $1/(1+\nu_c z) = 0.4667(57)$ in $d=2$ and $1/(1+\nu_c z) = 0.5602(69)$ in $d=3$. These estimates are shown as the blue band in the inset.

Using the literature values for the static critical exponents given in \cref{tab:crit_exponents}, we compute the dynamic critical exponent $z$ from our estimates of the scaling exponent $1/(1+\nu_c z)$ presented in \cref{fig:z_linregr}.
The estimates for the dynamic critical exponent $z$ we obtain are $z=2.142(49)$ in $d=2$ and $z=1.949(54)$ in $d=3$.
The results are summarized in \cref{tab:z} together with other literature values for comparison.
\begin{table}[tb] \centering
    \caption{\label{tab:z} Estimates for the dynamic critical exponent $z$ of Model~A in two and three dimensions from various other methods from the literature for comparison. These include previous estimates from classical-statistical lattice simulations of our model, Monte Carlo simulations, perturbative field theory, and non-perturbative renormalization group calculations as well as experimental results.
    }
    \begin{tabular}{l l l}
        \toprule
        Reference            & $d=2$                                         & $d=3$                              \\
        \midrule
        This work            & 2.142(49)                                     & 1.949(54)                          \\
        Class. stat.         & 2.10(4)~\cite{Schweitzer:2020noq}             & 1.92(11)~\cite{Schweitzer:2020noq} \\
        Monte Carlo          & 2.1667(5)~\cite{Nightingale:2000}             & 2.0245(15) ~\cite{Hasenbusch:2020} \\
        $\epsilon$ expansion & 2.14(2)~\cite{Adzhemyan:2021hvo}              & 2.0236(8)~\cite{Adzhemyan:2021hvo} \\
        Renorm. Group        & 2.15~\cite{Duclut:2017}                       & 2.024~\cite{Duclut:2017}           \\
        Experiment           & 2.09(6) (95\% confidence)~\cite{Dunlavy:2005} & 1.96(11)~\cite{Livet:2018}         \\
        \bottomrule
    \end{tabular}
\end{table}
Our result for the dynamic critical exponent in $d=2$ is in good agreement and compatible with all other present literature estimates.
This is not fully the case for the $d=3$ result, which is slightly lower than other theoretical estimates. Notably, Wilson's expansion methods predict $z=2+c\eta$ with $c \geq 0$ where $\eta$ is the anomalous dimension or static Fisher exponent~\cite{Halperin:1972bwo}. The agreement of our result with the experimental value in $d=3$ is therefore most likely coincidental as the uncertainties in the experimental result are also quite large.
The discrepancy between our result and the other theoretical estimates of slightly more than one standard deviation is most likely due to the finite size and quench rate limitations of our study. To obtain a more precise estimate of the dynamic critical exponent, one would need to broaden the dynamic scaling region by simulating larger system sizes which would allow us to go to smaller quench rates. This however would also require a significant increase in computational resources.
It might therefore be more feasible for future studies to try and obtain a heuristic parametrization of the corrections to the Kibble-Zurek scaling laws in fast quenches, which could then be used to extrapolate to the limit of zero quench rate. This would also eliminate a potential source of systematic error in the determination of the dynamic critical exponent due to the somewhat arbitrary definition of the dynamic scaling region.

\section{Numerical methods}

We integrate \cref{eq:eom1,eq:eom2} using a second-order Verlet-type integrator scheme, also known as ``leapfrog'' integration, evolving the system from time $t$ to $t+\Delta t$ in $N$ time steps $\delta t$, such that the evolution is stable and no step-size dependence is observed in the results. Typically, we performed measurements of observables in steps of $\Delta t = 0.1$ and used $N=16$ integrator steps per measurement, corresponding to a time step of $\delta t = 0.00625$.

The data used in this work was generated on the local heterogeneous cluster of the Giessen group consisting of mixed NVIDIA GPU hardware.
The code was written in {Julia}~\cite{Julia-2017} making use of the \verb|ParallelStencil.jl|~\cite{omlin2022highperformance} package for GPU acceleration. Visualizations and plots were created with \verb|Makie.jl|~\cite{Makie}.
The random number generation for the Gaussian noise $\eta_x$ was performed directly on the GPU using a counter-based Philox 2x32 random number generator provided by the \verb|CUDA.jl|~\cite{besard2018juliagpu} package. At the start of every simulation, the random number generator was uniquely seeded with the current UNIX epoch timestamp in milliseconds.

\bibliographystyle{elsarticle-num}
\bibliography{refs.bib}

\end{document}